# Genomic study of the Ket: a Paleo-Eskimo-related ethnic group with significant ancient North Eurasian ancestry

Pavel Flegontov[1,2,3]*, Piya Changmai[1,§], Anastassiya Zidkova[1,§], Maria D. Logacheva[2,4], Olga Flegontova[3], Mikhail S. Gelfand[2,4], Evgeny S. Gerasimov[2,4], Ekaterina E. Khrameeva[5,2], Olga P. Konovalova[4], Tatiana Neretina[4], Yuri V. Nikolsky[6,11], George Starostin[7,8], Vita V. Stepanova[5,2], Igor V. Travinsky[#], Martin Tříska[9], Petr Tříska[10], Tatiana V. Tatarinova[2,9,12]*

[1] *Department of Biology and Ecology, Faculty of Science, University of Ostrava, Ostrava, Czech Republic*

[2] *A.A.Kharkevich Institute for Information Transmission Problems, Russian Academy of Sciences, Moscow, Russian Federation*

[3] *Institute of Parasitology, Biology Centre, Czech Academy of Sciences, České Budějovice, Czech Republic*

[4] *Department of Bioengineering and Bioinformatics, Lomonosov Moscow State University, Moscow, Russian Federation*

[5] *Skolkovo Institute of Science and Technology, Skolkovo, Russian Federation*

[6] *Biomedical Cluster, Skolkovo Foundation, Skolkovo, Russian Federation*

[7] *Russian State University for the Humanities, Moscow, Russian Federation*

[8] *Russian Presidential Academy (RANEPA), Moscow, Russian Federation*

[9] *Children's Hospital Los Angeles, Los Angeles, CA, USA*

[10] *Instituto de Patologia e Imunologia Molecular da Universidade do Porto (IPATIMUP), Porto, Portugal*

[11] *George Mason University, Fairfax, VA, USA*

[12] *Spatial Science Institute, University of Southern California, Los Angeles, CA, USA*

*corresponding authors: P.F., email pavel.flegontov@osu.cz; T.V.T., email tatiana.tatarinova@usc.edu

[§] the authors contributed equally

[#] retired, former affiliation: Central Siberian National Nature Reserve, Bor, Krasnoyarsk Krai, Russian Federation.

## Abstract

The Kets, an ethnic group in the Yenisei River basin, Russia, are considered the last nomadic hunter-gatherers of Siberia, and Ket language has no transparent affiliation with any language family. We investigated connections between the Kets and Siberian and North American populations, with emphasis on the Mal'ta and Paleo-Eskimo ancient genomes using original data from 46 unrelated samples of Kets and 42 samples of their neighboring ethnic groups (Uralic-speaking Nganasans, Enets, and Selkups). We genotyped over 130,000 autosomal SNPs, determined mitochondrial and Y-chromosomal haplogroups, and performed high-coverage genome sequencing of two Ket individuals. We established that the Kets belong to the cluster of Siberian populations related to Paleo-Eskimos. Unlike other members of this cluster (Nganasans, Ulchi, Yukaghirs, and Evens), Kets and closely related Selkups have a high degree of Mal'ta ancestry. Implications of these findings for the linguistic hypothesis uniting Ket and Na-Dene languages into a language macrofamily are discussed.

## Key words:

Migration, genetics, genome sequencing, Ket, Na-Dene languages, Yenisei River

## Introduction

The Kets (an ethnic group in the Yenisei River basin, Russia) are among the least studied native Siberians. Ket language lacks transparent affiliation with any major language family, and is clearly distinct from surrounding Uralic, Turkic and Tungusic languages(Vajda 2004). Moreover, until their forced settlement in 1930s, Kets were considered the last nomadic hunter-gatherers of North Asia outside the Pacific Rim(Vajda 2009).

Ket language, albeit almost extinct, is the only language of the Yeniseian family that survived into the 21st century. According to toponymic evidence, prior to the 17th century speakers of this language family occupied vast territories of Western and Central Siberia, from northern Mongolia in the south to the middle Yenisei River in the north and from the Irtysh River in the west to the Angara River in the east(Dul'zon 1959, Vajda 2001). Most Yeniseian-speaking tribes used to live south of the current Ket settlements. Ancestors of the Yeniseian people were tentatively associated(Chlenova 1975) with the Karasuk Culture (3200-2700 YBP) of the upper Yenisei(Allentoft, Sikora et al. 2015). Over centuries, Kets and other Yeniseian people suffered relocation, extinction and loss of language and culture. First, they were under a constant pressure from the reindeer herders to the north (Enets and Nenets) and east (Evenks) and the Turkic-speaking pastoralists to the south. Second, Russian conquest of Siberia, which started at the end of the 16th century, exposed the natives to new diseases, such as the 17th century smallpox epidemic(Alekseenko 1967). Third, in the 20th century USSR resettled the Kets in Russian-style villages, thus interrupting their nomadic life-style(Vajda 2009). Almost all Yeniseian-speaking tribes (Arin, Assan, Baikot, Pumpokol, Yarin, Yastin) have disappeared by now. Under pressure of disease and conflict, the Kets have been gradually migrating north along the Yenisei River, and now reside in several villages in the Turukhansk district (Krasnoyarsk region); around 1,200 people in total.(Krivonogov 2003)

Yeniseian linguistic substrate is evident in many contemporary Turkic languages of South Siberia: Altaian, Khakas, Shor, Tubalar, Tuvinian, and in Mongolic Buryat language. As these languages are spoken in river basins with Yeniseian river names(Vajda 2004), the Yeniseian tribes were likely to have mixed with these ethnic groups (and with the Southern Samoyedic groups Kamasins and Mators, now extinct(Vajda 2004)) at different times. We expect to find genetic signatures of these events.

Until the 20th century, Kets, being nomadic hunters and fishers in a vast Siberian boreal forest, had little contact with other ethnic groups, which is manifested by the paucity of loanwords in Ket language(Vajda 2009). However, since the collapse of the exogamous marriage system following smallpox epidemics in the 17th and 18th centuries, the Kets have been marrying Selkups, Uralic-speaking reindeer herders(Vajda 2009, Vajda 2011). Moreover, during the 20th century, the settled Kets have been increasingly mixing with the Russians and native Siberian people, which resulted in irrevocable loss of Ket language, genotype, and culture.

Recently, a tentative link was proposed between the Yeniseian language family and the Na-Dene family of Northwest North America (composed of Tlingit, Eyak, and numerous Athabaskan languages), thus forming a Dene-Yeniseian macrofamily.(Comrie 2010, Vajda 2011) The Dene-Yeniseian-linkage is viewed by some as the first relatively reliable trans-Beringian language connection(Comrie 2010), with important implications on timing of the alleged Dene-Yeniseian population split, the direction of the subsequent migration (from or to America), the possible language shifts and population admixture(Ives 2010, Potter 2010, Scott and O'Rourke 2010).

So far, no large-scale population study was conducted with samples from each of the presently occupied Ket villages. Previously, six Ket individuals were genotyped(Rasmussen, Li et al. 2010, Elhaik, Greenspan et al. 2013, Fedorova, Reidla et al. 2013, Raghavan, Skoglund et al. 2014, Seguin-Orlando, Korneliussen et al. 2014, Raghavan, Steinrucken et al. 2015) and two of them sequenced.(Raghavan, Steinrucken et al. 2015) These studies concluded that the Kets do not differ from surrounding Siberian populations, which is rather surprising, given their unique language and ancient hunter-gatherer life-style. In order to clarify this issue, in 2013 and 2014, we collected 57 (46 unrelated) samples of Kets and 42 unrelated samples of their neighboring Uralic-speaking ethnic groups (Nganasans inhabiting the Taymyr Peninsula, and Enets and Selkups living further south along the Yenisei). We genotyped approximately 130,000 autosomal SNPs and determined mitochondrial and Y-chromosomal haplogroups with the GenoChip array.(Elhaik, Greenspan et al. 2013) We also performed high-coverage genome sequencing of two Ket individuals. Using these data, we investigated connections between Kets and several modern and ancient Siberian and North American populations (including the Mal'ta and Saqqaq ancient genomes). In addition, we estimated Neanderthal contribution in Kets' genome and in specific gene groups.

Mal'ta is a ~24,000 YBP old Siberian genome, recently described(Raghavan, Skoglund et al. 2014) as a representative of ancient North Eurasians, ANE(Lazaridis, Patterson et al. 2014), a previously unknown northeastern branch of the Eurasian Paleolithic population. ANE contributed roughly 30% of the gene pool of Native Americans of the first settlement wave(Raghavan, Skoglund et al. 2014) and reshaped the genetic landscape of Central and Western Europe in the Bronze Age around 5,000-4,000 YBP, when ANE genetic pool was introduced into Europe via expansion of the Corded Ware culture(Allentoft, Sikora et al. 2015, Haak, Lazaridis et al. 2015).

A global maximum of ANE ancestry occurs in Native Americans, with lower levels in peoples of more recent Beringian origin, i.e. indigenous populations of Chukotka, Kamchatka, the Aleutian Islands and the American Arctic(Lazaridis, Patterson et al. 2014, Raghavan, DeGiorgio et al. 2014, Raghavan, Skoglund et al. 2014). In modern Europe, ANE genetic contribution is the highest in the Baltic region, on the East European Plain and in the North Caucasus(Lazaridis, Patterson et al. 2014, Allentoft, Sikora et al. 2015, Haak, Lazaridis et al. 2015). However, little is known about the distribution of ANE ancestry in its Siberian homeland. According to a single $f_4$ statistic, the Kets had the third highest value of ANE genetic contribution among all Siberian ethnic groups, preceded only by Chukchi and Koryaks(Seguin-Orlando, Korneliussen et al. 2014). Thus, we suggest that the Kets might represent the peak of ANE ancestry in Siberia; the hypothesis we tested extensively in this study.

Saqqaq genome (~4,000 YBP) from Greenland(Rasmussen, Li et al. 2010) represents the Saqqaq archeological culture (4,500-2,800 YBP). This culture forms a continuum with Dorset and Norton cultures (2,500-1,000 YBP). Together, they are termed Paleo-Eskimo.(Raghavan, DeGiorgio et al. 2014) Paleo-Eskimos were culturally and genetically distinct from modern Inuits and Eskimos(Potter 2010, Raghavan, DeGiorgio et al. 2014). The Saqqaq culture is part of the wider Arctic Small Tool tradition (ASTt) that had rapidly spread across Beringia and the American Arctic coastal (but not the interior) regions after 4,800 YBP, bringing pottery, bow and arrow technology to the northern North America(McGhee 1996, Ives 2010, Potter 2010). According to the archaeological data, the likely source of this spread was located in Siberia, namely in the Lena River basin (probably, the Bel'kachi culture(Potter 2010)). On genetic grounds, Paleo-Eskimos were also argued to represent a separate migration into America(Rasmussen, Li et al. 2010, Reich, Patterson et al. 2012, Raghavan, DeGiorgio et al. 2014). ASTt spread coincided the arrival of mitochondrial haplogroup D2 into America and the spread of haplogroup D2a(Tamm, Kivisild et al. 2007); the Saqqaq individual bore haplogroup D2a1(Gilbert, Kivisild et al. 2008). The closest modern relatives of Saqqaq occur among Beringian populations (Chukchi, Koryaks, Inuits(Raghavan, DeGiorgio et al. 2014)) and Siberian Nganasans(Rasmussen, Li et al. 2010). In addition, Saqqaq has been linked to Na-Dene-speaking Chipewyans (16% contribution to this population modeled with admixture graphs(Reich, Patterson et al. 2012)). However, mitochondrial haplogroup data(Hayes, Coltrain et al. 2002, Gilbert, Kivisild et al. 2008, Raghavan, DeGiorgio et al. 2014) argues against the proximity of Paleo-Eskimos to contemporary Na-Dene people(Ives 2010, Potter 2010), primarily due to the very high frequency of haplogroup A in the latter(Malhi, Mortensen et al. 2003) (Suppl. Information Section 10). Archeological evidence seems to support this argument(Potter 2010).

There is no archaeological evidence of considerable trans-Beringian population movements between the inundation of the Bering Platform around 13,000-11,000 YBP and 4,800 YBP. Therefore, it is unlikely that the hypothetical Dene-Yeniseian language family has separated prior to 11,000 YBP, according to current concepts of language evolution(Ives 2010, Potter 2010). ASTt could be the vehicle spreading Dene-Yeniseian languages and genes from Siberia to Alaska and to the American Arctic(Potter 2010). However, as argued based on language phylogenetic trees(Sicoli and Holton 2014) in the framework of the Beringian standstill model,(Tamm, Kivisild et al. 2007, Pringle 2014) the Dene-Yeniseian languages have originated in Beringia and spread in both directions. Irrespective of the migration direction and their relationship to contemporary Na-Dene groups, Paleo-Eskimos are the primary target for investigating genetic relationship with the Kets.

In this study, we claim the following: (1) Kets and Selkups form a clade closely related to Nganasans; (2) Nganasans, Kets, Selkups, Ulchi, Yukaghirs, and possibly Evens form a group of populations related to Paleo-Eskimos; (3) unlike the other members of this group, Kets (and Selkups to a lesser extent) derive roughly 30-60% of their ancestry from ancient North Eurasians, and represent the peak of ancient North Eurasian ancestry among all investigated modern Eurasian populations west of Chukotka and Kamchatka.

## Results and Discussion

*Identification of a non-admixed Ket genotype*

We compared the GenoChip SNP array data for the Kets, Selkups, Nganasans, and Enets populations (Suppl. file S1) to the worldwide collection of populations(Elhaik, Tatarinova et al. 2014) based on 130K ancestry-informative markers.(Elhaik, Greenspan et al. 2013) We applied GPS(Elhaik, Tatarinova et al. 2014) and reAdmix(Kozlov, Chebotarov et al. 2015) algorithms to infer provenance of the samples and confirm self-reported ethnic origin. According to the GPS analysis, 46 of 57 (80%) self-reported Kets were identified as Kets, 9 (16%) as Selkups, one as a Khakas, and one as a Dolgan (Turkic speakers from the Taymyr Peninsula). In addition to the proposed population and geographic location, GPS also reports prediction uncertainty (the smallest distance to the nearest reference population) (Suppl. Fig. 4.2). The average prediction uncertainty was 2.5% for individuals identified as Kets; 4.4% for Selkpus; 5.6% for Khakas, and 3.9% for the Dolgan. Prediction uncertainty over 4% indicates that the individual is of a mixed origin and the GPS algorithm is not applicable.

Using the reAdmix approach (in the unconditional mode)(Kozlov, Chebotarov et al. 2015), we represented 57 Kets as weighted sums of modern reference populations (Suppl. Fig. 4.3, Suppl. Table 3). The median weight of the Ket ancestry in self-identified Kets was 94%; 39 (68%) of them had over 90% of the Ket ancestry (non-admixed Kets). Seven individuals with self-reported purely Ket origin appear to be closer to Selkups, with median 89% percent of Selkup ancestry. This closeness is not surprising, given the long shared history of Ket and Selkup people(Vajda 2004). Individuals with incorrect self-identification were randomly distributed across sixteen birthplaces along the Yenisei River (Suppl. Table 3). The identity by descent (IBD) analysis shows presence of clusters consistent with self-identification of ethnic groups, and supports the genetic proximity of Kets and Selkups (Suppl. Fig. 4.4)

86% of GPS predictions agree with the major ancestry prediction by reAdmix (Suppl. Table 3). The Pearson's correlation between percentage of major ancestry and GPS uncertainty is -0.42, meaning that the individuals predicted by reAdmix to be of non-admixed origin are likely to be predicted to be non-admixed by GPS as well. Hence, we identified a subset of non-admixed Kets among self-identified Ket individuals, and nominated individuals for whole-genome sequencing.

*'Ket-Uralic' admixture component*

The National Genographic dataset(Elhaik, Tatarinova et al. 2014) included only a few Siberian populations. Hence, following the exclusion of first-, second-, and third-degree relatives among the individuals genotyped in this study (Suppl. file S1, Suppl. Fig. 4.1), we combined the GenoChip array data with published SNP array datasets to produce a worldwide dataset of 90 populations and 1,624 individuals (Methods; Suppl. Table 1). The intersection dataset, containing 32,189 SNPs (Suppl. Table 1), was analyzed with ADMIXTURE(Alexander, Novembre et al. 2009) (Fig. 1), selecting the best of 100 iterations and using the 10-fold cross-validation criterion. At K≥11, ADMIXTURE identified a characteristic component for the Ket population (Suppl. Information Section 5). This component reached its global maximum of nearly 100% in Kets, closely followed by Selkups from this study (up to 81.5% at K=19), reference Selkups (up to 48.5%) and Enets (up to 22.6%). The difference between the Selkups from this study and the reference Selkups(Raghavan, Skoglund et al. 2014) can be attributed to a much closer geographic proximity of the former population to the settlements of Kets, with whom they have a long history of cohabitation and mixture with(Vajda 2009, Vajda 2011).

The 'Ket' component occurred at high levels (up to ~20%) in four populations of the Altai region: Shors, Khakases, Altaians, and Teleuts, all of them Turkic-speaking. The Altai region was populated by Yeniseian-speaking people before they were forced to retreat north (Suppl. Information section 2), and, therefore, our results agree with earlier suggestions of the ethnic mixture history in Siberiua. Lower levels of the 'Ket' component, from 15% to 5%, were observed in the following geographic regions (in decreasing order): Volga-Ural region (in Mari, Chuvashes, Tatars, Mordovians); Central and South Asia (Kazakhs, Kyrgyz, Uzbeks, Tajiks, Turkmens, Indians); East Siberia and Mongolia (Yakuts, Dolgans, Evenks, Buryats, Mongols, Nganasans from this study, Evens); North Caucasus (Nogays, Ingushes, Balkars). The 'Ket' component also occurred at a low level in Russians (up to 7.1%), Finns (up to 5.4%), and, remarkably, in the Saqqaq ancient genome from Greenland (7.2%, see below). These results are further supported by the principal component analysis, PCA (Suppl. Information Section 6).

In order to verify and explain the geographic distribution of the 'Ket' admixture component, we have performed ADMIXTURE analysis on three additional datasets, varying in populations (Suppl. Table 2) and marker selection (Suppl. Table 1) (see Suppl. Information Section 5). The analysis suggested the existence of an admixture component with a peculiar geographic distribution, not discussed in previous studies. This component is characteristic not only of Kets, but also of Samoyedic-speaking and Ugric-speaking people of the Uralic language family: Selkups, Enets, Nenets, Khanty, Mansi, with a notable exception of Samoyedic-speaking Nganasans. Association of this component with Uralic-speaking ethnic groups may explain its appearance in the Volga-Ural region, where Finnic-speaking Mari and Mordovians reside alongside Chuvashes and Tatars, Turkic-speaking groups with a notable Uralic linguistic substrate(Johanson 2010). All of the above populations feature moderate levels of the Ket-Uralic component, with maximum values encountered in a given population ranging from 5.5% to 22% in different datasets (Suppl. Table 4). Lower levels of this component (5.3-8.6%) are observed in Finns and Russians, the latter known to be mixed with Uralic-speaking people in historic times(Flegontova, Khrunin et al. 2009, Khrunin, Khokhrin et al. 2013). High levels of the Ket-Uralic admixture component in South Siberia are in agreement with the former presence of extinct Yeniseian- and Samoyedic-speaking ethnic groups there(Vajda 2004). Moderate levels of the Ket-Uralic component in Central and South Asia might be due to hypothetical nomadic tribes, representing steppe offshoots of Yeniseian- or Uralic-speaking hunter-gatherers of the forest zone. According to some interpretations(Vovin 2000, Vovin 2002), Jie, one of five major tribes of the Xiongnu nomadic confederation occupying northern China and Mongolia in the 3$^{rd}$ century BC – 2$^{nd}$ century AD, spoke a Yeniseian language possibly close to Pumpokol, the southernmost extinct Yeniseian language stretching into Mongolia(Vajda 2009). However, the most intriguing is the appearance of the Ket-Uralic component in the Saqqaq

Paleo-Eskimo (~4,000 YBP): at a low level of 6.3-8.6%, but consistently in all three datasets containing this individual (Suppl. Table 4).

In summary, we demonstrated existence of the admixture component specific for Kets and Uralic-speaking populations, featuring peculiar geographic distribution.

*Kets in the context of Siberian populations*

The Ket and Selkup populations were closely related according to multiple analyses (see PCA plots in Suppl. Figs. 6.3, 6.8, and $f_3$ and $f_4$ statistics(Patterson, Moorjani et al. 2012) in Suppl. Information Sections 7 and 8) and formed a clade with Nganasans (Fig. 2A). Nganasans appeared as the closest relatives of both populations according to statistics $f_3$(Yoruba; Ket, X) (Suppl. Figs. 7.1-7.7), $f_3$(Yoruba; Selkup, X) (Suppl. Figs. 7.11-7.16) and $f_4$(Ket, Chimp; Y, X) (Suppl. files S3, S4) computed on various datasets (Suppl. Table 1). Statistic $f_3$(O; A, $X_1$)(Patterson, Moorjani et al. 2012) measures relative amount of genetic drift shared between the test population A and a reference population $X_1$, given an outgroup population O, distant from A and $X_1$. Statistic $f_4$(X, O; A, B)(Patterson, Moorjani et al. 2012) tests whether A and B are equidistant from X, given a sufficiently distant outgroup O: in that case the statistic is close to zero. Otherwise, the statistic shows whether X is more closely related to A or to B.

Nganasans were consistently scored as the top or one of five top hits for Kets, in addition to Selkups, Yukaghirs, and Beringian populations (Suppl. Figs. 7.1-7.7, Suppl. files S3, S4); and Yukaghirs, Evenks, Ulchi, and Dolgans were recovered as top hits for Nganasans in different datasets (Suppl. Figs. 7.17-7.22). Nganasans, Ulchi, and Yukaghirs appeared as the closest Siberian relatives of the Saqqaq Paleo-Eskimo (not counting the populations of Chukotka and Kamchatka, e.g., Chukchi, Eskimos, Itelmens, and Koryaks), according to statistics $f_3$(Yoruba; Saqqaq, X) (Suppl. Figs. 7.47, 7.48) and $f_4$(Saqqaq, Chimp; Y, X) (Suppl. file S5) and in agreement with previous results.(Rasmussen, Li et al. 2010, Raghavan, DeGiorgio et al. 2014)

In line with these results, Nganasans, Kets, Selkups, Evens, and Yukaghirs formed a clade in a maximum likelihood tree constructed with TreeMix on a HumanOrigins-based dataset of 194,750 SNPs. The migration edge appeared between the Saqqaq branch and the base of this clade, showing 34% of Siberian ancestry in Saqqaq (Fig. 2A,B). TreeMix analysis predicted 37-59% Ket ancestry in the Saqqaq and Late Dorset Paleo-Eskimo genomes on a larger genome-based dataset of 347,466 SNPs (Fig. 2C,D) and its version without transitions (185,382 SNPs, Suppl. Figs. 9.20, 9.21). As the dataset lacked Nganasan or Yukaghir genomes (not available at the time of study), Kets were the only representative of the Nganasan-related Siberian clade in this dataset. In line with this result, Saqqaq and Late Dorset appeared as the top hits for Kets, followed by Native American groups, according to statistic $f_3$(Yoruba; Ket, X) applied on the full-genome dataset (Suppl. Fig. 7.8). These results were reproduced with $f_3$(Yoruba; Ket, X) and $f_4$(Ket, Yoruba; Y, X) on the dataset without transitions, using both Ket genomes (Ket884 and Ket891) or only Ket891 (Fig. 3B,C, Suppl. Figs. 7.9, 7.10, 8.37). In addition, all possible population pairs (X, Y) were tested with $f_4$(Saqqaq, Yoruba; Y, X) on the full-genome dataset. Compared to Kets, Saqqaq was significantly closer only to Greenlanders (Z-score of -2.9) and Late Dorset (Z-score of -13.9) (Suppl. Fig. 8.40A). The respective Z-scores on the dataset without transitions were -2 and -11.7 (Suppl. Fig. 8.40B).

In our ADMIXTURE analyses on all datasets (Fig. 1A, Suppl. Figs. 5.4 and 5.7), the Saqqaq individual featured the following components: Eskimo (Beringian), Siberian, and South-East Asian. This order is in perfect agreement with the original study of the Saqqaq genome.(Rasmussen, Li et al. 2010) Although the Ket-Uralic component was low in Saqqaq (6.3-8.6%, Fig. 1C, Suppl. Figs. 5.6 and 5.9), it appeared in all analyzed datasets. Moreover, PC3 vs. PC4 plots for two HumanOrigins-based datasets placed Saqqaq close to Ket, Selkup, Mansi, Yukaghir, and Even individuals (Fig. 4B, Suppl. Fig. 6.10). Three former populations showed considerable levels of the Ket-Uralic admixture component (>14%, see Suppl. Table 4). These analyses also support the fact that Kets belong to a cluster of Siberian populations most closely related to Saqqaq.

Accepting the model that Saqqaq represents a mixture of Beringian and Siberian populations (e.g., see the Ket-Saqqaq and Greenlander-Saqqaq migration edges in Fig. 2C), and the tree topology in which Native American and Beringian populations form a clade relative to Kets, Nganasans, and Yukaghirs (Fig. 2A, Suppl. Information Section 8), we can estimate the percentage of Siberian ancestry in Saqqaq using $f_4$-ratios(Patterson, Moorjani et al. 2012):

$$1 - \frac{f4(\text{Karitiana,Outgroup; Saqqaq,Ket})}{f4(\text{Karitiana,Outgroup; Beringian population,Ket})}.$$

According to this method, the Siberian ancestry in Saqqaq ranged from 63% to 67%, using various outgroups in the genome-based dataset without transitions (Suppl. file S7). A similar estimate, 59%, was obtained by TreeMix on the original genome-based dataset (Fig. 2C).

In summary, we conclude that Kets and Selkups belong to a group of Siberian populations most closely related to ancient Paleo-Eskimos, represented by the Saqqaq genome.

*Mal'ta (ancient North Eurasian) ancestry in Kets*

Unlike the other members of the Nganasan-related clade (Fig. 2A), Kets and, to a lesser extent, Selkups have a high proportion of Mal'ta ancestry, alternatively referred to as ancient North Eurasian ancestry(Lazaridis, Patterson et al. 2014). As calculated by statistic $f_3$(Yoruba; Mal'ta, X) on the full-genome dataset, Ket891 is placed in the gradient of genetic drift shared with Mal'ta, ahead of all Native Americans of the first settlement wave and second after Motala12 (Fig. 5), an approximately 8,000 year old hunter-gatherer genome from Sweden(Lazaridis, Patterson et al. 2014). Notably, ancient North Eurasian ancestry in Motala12 was estimated at ~22%(Lazaridis, Patterson et al. 2014, Haak, Lazaridis et al. 2015). This fact may explain that Motala12 is the best hit for Mal'ta in our $f_3$ statistic set-up. In the full-genome dataset without transitions (main source of ancient DNA biases(Axelsson, Willerslev et al. 2008)), the Ket891 genome was the fourth best hit for Mal'ta, after Motala12, Karitiana, and Mixe (Suppl. Fig. 7.42). Also, the Kets were consistently placed at the top of the Eurasian spectrum of $f_3$(Yoruba; Mal'ta, X) values (Suppl. Fig. 7.36, 7.37) or within the American spectrum (Suppl. Figs. 7.38-39) by statistics $f_3$(Yoruba; Mal'ta, Ket891) and $f_3$(Yoruba; Mal'ta, Ket884+891) computed for two datasets combining the Ket genomes and SNP array data (Suppl. Table 1).

These results were consistent with calculations of $f_4$ statistic in two configurations: (X, Chimp; Mal'ta, Stuttgart) or (X, Papuan; Sardinian, Mal'ta), reproducing the previously used statistics(Lazaridis, Patterson et al. 2014, Seguin-Orlando, Korneliussen et al. 2014) (Suppl. Figs. 8.1-8.8, 8.25-8.32). $f_4$(X, Chimp; Mal'ta, Stuttgart) analysis tests whether the population X has more drift shared with Mal'ta or with Stuttgart (an early European farmer, EEF(Lazaridis, Patterson et al. 2014)). Sardinians were used as the closest modern proxy for EEF(Lazaridis, Patterson et al. 2014) in $f_4$(X, Papuan; Sardinian, Mal'ta). All possible population pairs (X,Y) were tested by $f_4$(Mal'ta, Yoruba; Y, X) on the full-genome dataset including both Ket individuals (Fig. 3A). Compared to Kets, Mal'ta was probably closer only to Motala12, although with a non-significant Z-score of -1.1. As expected, the results changed with the individual Ket884 removed: Z-score for $f_4$(Mal'ta, Yoruba; Ket891, Motala12) became even less significant, -0.4 (Fig. 3A). Mal'ta ancestry in Kets was further supported by the TreeMix(Pickrell and Pritchard 2012) analysis (Fig. 2C).

Based on all analyses, we can tentatively model Kets as a two-way mixture of East Asians and ancient North Eurasians (ANE). Therefore, ANE ancestry in Kets can be estimated using various $f_4$-ratios (see details in Suppl. Information Section 8) from 27% to 62% (depending on the dataset and reference populations), vs. 2% in Nganasans, 30 – 39% in Karitiana, and 23 – 28% in Mayans (Suppl. file S7). Integrating data by different methods, we conservatively estimate that Kets have the highest degree of ANE ancestry among all investigated modern Eurasian populations west of Chukotka and Kamchatka. We speculate that ANE ancestry in Kets was acquired in the Altai region, where the Bronze Age Okunevo culture was located, with a surprisingly close association with Mal'ta(Allentoft, Sikora et al. 2015). Later, Yeniseian-speaking people occupied this region until the 16$^{th}$-18$^{th}$ centuries(Dul'zon 1959, Vajda 2001). We suggest that Mal'ta ancestry was introduced into Uralic-speaking Selkups later, starting to mix with Kets extensively in the 17-18$^{th}$ centuries(Vajda 2009, Vajda 2011).

*Kets and Na-Dene speakers*

In this study, Na-Dene-speaking people were represented by Athabaskans, Chipewyans, Tlingit, and, possibly, Haida. The latter language was originally included into the Na-Dene language family(Durr and Renner 1995), although this affiliation is now disputed.(Vajda 2011) Na-Dene-speaking people were suggested to be related, at least linguistically, to Yeniseian-speaking Kets(Vajda 2011). ADMIXTURE, PCA, $f_3$ and $f_4$ statistics, and TreeMix analyses do not suggest any specific link between Kets and Athabaskans, Chipewyans, or Tlingit (see Suppl. Information section 8). TreeMix constructed tree topologies where Athabaskans or Chipewyans formed a stable highly supported clade with other Native Americans (bootstrap support from 88 to 98, Fig. 2A,C). This topology was supported by statistics $f_4$ (Athabaskan, Yoruba; Ket, X), which demonstrated significantly negative Z-scores, < -5, for Clovis, Greenlanders, Karitiana, Mayans, and Mixe (Suppl. Fig. 8.42) on the genome-based dataset with or without transitions. Notably, the same topology was previously demonstrated for Athabaskans.(Raghavan, Steinrucken et al. 2015) Non-significant Z-scores < -2 for the statistic $f_4$(Athabaskan, Yoruba; Ket, Saqqaq/Late Dorset) are consistent with Kets and Paleo-Eskimos forming a clade distinct from Athabaskans (see TreeMix results in Suppl. Figs. 9.7-9.10).

Note, that the Arctic Small Tool tradition the Saqqaq culture belongs to, may reflect the Dene-Yeniseian movement over the Bering Strait(Ives 2010, Potter 2010). According to the admixture graph analysis modeling relationships among Chipewyans, Saqqaq, Algonquin, Karitiana, Zapotec, Han, and Yoruba(Reich, Patterson et al. 2012), only one topology fits these data, in which Chipewyans represent a mixture of 84% First Americans and 16% Saqqaq, and Saqqaq a mixture of 16% First Americans and 84% of an Asian source distinct from that of First Americans(Reich, Patterson et al. 2012). Notably, our estimates using $f_4$-ratios are similar: 18-27% Saqqaq ancestry in Chipewyans and 10-15% in Athabaskans (Suppl. file S7). Considering 67% as the highest proportion of Siberian ancestry in Saqqaq obtained in this study, Chipewyans feature ~10.7% of Siberian ancestry, i.e. 67%×16%. Similarly, only 1.4% (noise level) of the Ket-Uralic admixture component is predicted in Chipewyans, with 8.6% as the highest percentage of this component found in Saqqaq (Suppl. Fig. 5.9). Therefore we cannot reliably support the hypothetical genetic connection between Yeniseian and Na-Dene-speaking people, with the latter undergoing massive admixture with the First Americans, provided the employed methods and population samples.

However, a weak signal was detected with the outgroup $f_3$ statistic $f_3$(Yoruba; Haida, X) on the HumanOrigins-based dataset (Fig. 6, Suppl. Fig. 7.32): Kets emerged as the best hit to Haida in Eurasia, west of Chukotko-Kamchatkan (Beringian) populations, whereas Nganasans are the best hit to Chipewyans and Tlingit according to outgroup $f_3$ statistic (Suppl. Figs. 7.31, 7.33). However, $f_4$(Haida, Chimp; Ket, X) produced Z-scores for a number of Eurasian populations, e.g. Nganasans and Saami, close to zero (0.14 and 0.11, respectively, Suppl. Fig. 8.33), in line with only a marginal difference in statistics $f_3$(Yoruba; Haida, Ket) and $f_3$(Yoruba; Haida, Nganasan) (Suppl. Fig. 7.32). The difference with the second best hit, Nganasans, is more prominent in the dataset without the purportedly mixed individual Ket884 (Fig. 6). Remarkably, in contrast to Chipewyans and Athabaskans, Haida and Tlingit represent coastal populations that might have came into close contact with (or have been a part of) the mainly coastal ASTt archaeological culture(Potter 2010).

Hopefully, the question of the Dene-Yeniseian genetic relationship and its correlation with the linguistic relationship, will be answered definitely with a study of complete genomes of Athabaskans, Chipewyans, Tlingit and Haida, combined with those of Kets, Nganasans and other relevant reference groups.

*Mitochondrial and Y-chromosomal haplogroups in Kets*

We have determined mitochondrial and Y-chromosomal haplogroups based on SNP data from GenoChip: approximately 3,300 mitochondrial and 12,000 Y-chromosomal SNPs (see Methods for details). The frequencies of mitochondrial haplogroups in 46 putatively unrelated Kets in this study (Suppl. file S8) were similar to those reported previously for 38 Ket individuals(Derbeneva, Starikovskaia et al. 2002). Notably, the frequency of mitochondrial haplogroup U4, predominant in Kets, correlated with proportion of the Ket-Uralic admixture component: Pearson's correlation coefficient reached up to 0.81 ($p$-value $5\times10^{-10}$) on three datasets analyzed (Suppl. Information Section 10). The Ket-Uralic admixture component did not significantly correlate with any other major mitochondrial haplogroup on two of three datasets analyzed (Suppl. files S9-11), and in the GenoChip-based dataset, the correlation of haplogroup U4 and the Ket-Uralic admixture component was associated with the second lowest $p$-value among all possible pairs of haplogroups and admixture components (Suppl. Table 5). Remarkably, ancient European hunter-gatherers had haplogroup U with >80% frequency(Bramanti, Thomas et al. 2009, Malmstrom, Gilbert et al. 2009, Fu, Meyer et al. 2013), and the Mal'ta individual also belonged to a basal branch of haplogroup U without affiliation to known subclades.(Raghavan, Skoglund et al. 2014) Therefore, haplogroup U, especially its U4 and U5 branches(Brandt, Haak et al. 2013), may be considered as a marker of West European hunter-gatherers (WHG) and of ancient North Eurasian (ANE) populations. In this light, high prevalence of haplogroup U4 in Kets and Selkups (Suppl. file S8) correlates well with large degrees of ANE ancestry in these populations.

The frequencies of Y-chromosomal haplogroups in 20 Ket males in this study were also similar to those reported previously for 48 Ket individuals(Tambets, Rootsi et al. 2004): more than 90% of Kets had haplogroup Q1a (of subclade Q1a2a1 as shown in our study) (Suppl. file S12), while haplogroups I1a2 and I2a1b3a occurred in just two Ket individuals. For Eurasian populations, frequency of haplogroup Q correlated with proportion of the Ket-Uralic admixture component: correlation coefficient reached up to 0.93 ($p$-value $1.7\times10^{-15}$) on three datasets analyzed (Suppl. Information section 10). The other major Y-chromosomal haplogroups demonstrated weaker correlation with the Ket-Uralic admixture component (data not shown). The Mal'ta individual had a Y-haplogroup classified as a branch basal to the modern R haplogroup, and the modern haplogroup Q formed another sister-branch of haplogroup R(Raghavan, Skoglund et al. 2014). It is tempting to hypothesize that haplogroup Q1a correlates with ANE ancestry on a global scale: both reach their maxima in America and in few Siberian populations including Kets. Moreover, haplogroup Q1a has been found in 1 out of 4 male individuals of the Bronze Age Karasuk archaeological culture (3,400-2,900 YBP), and in 2 out of 3 Iron Age individuals from the Altai region (3,000-1,400 YBP)[6]. Importantly, the Karasuk culture has been tentatively associated with Yeniseian-speaking people(Chlenova 1975), and the Altai region is covered by hydronyms of Yeniseian origin(Dul'zon 1959, Vajda 2001). Altai’s modern populations, as demonstrated in this study, have a rather large proportion of the Ket-Uralic admixture component.

*Neanderthal contribution in the Ket genomes*

To estimate the Neanderthal gene flow influence, we performed D-statistic analysis as described in Green et al.(Green, Krause et al. 2010) Given two Ket and two Yoruba individuals, we calculated the statistic *D*(Neanderthal, Chimp; Ket, Yoruba) for four different pairs of individuals. The mean *D*-statistic value, 3.85±0.15%, was in good agreement with other studies(Green, Krause et al. 2010, Khrameeva, Bozek et al. 2014). As a control, we replaced the Ket genotypes with Vietnamese genotypes processed using the same procedure. The control *D*-statistic value was 3.95±0.19% (Suppl. Table 6). Positive *D*-statistic values reflect higher similarity of Ket rather than Yoruba genotypes to Neanderthal genotypes, as expected for any non-African individuals. In order to find Ket functional gene groups enriched in Neanderthal alleles we applied the GSEA algorithm to 'biological process' GO terms(Khrameeva, Bozek et al. 2014) (Suppl. Information Section 11, Suppl. file S13). The only gene group significantly enriched in Neanderthal-like sites was 'amino acid catabolic process', with some genes involved in the urea cycle and in tyrosine and phenylalanine catabolism demonstrating the highest values of D-statistic (Suppl. Information Section 11). This finding may be explained by protein-rich meat diet characteristic of both Neanderthals(Sistiaga, Mallol et al. 2014) and Kets. We suggest that Kets, who abandoned the nomadic hunting lifestyle only in the middle of the $20^{th}$ century, are a good model of genetic adaptation to protein-rich diets.

## Conclusions and Outlook

Based on our results and previous studies,(Rasmussen, Li et al. 2010, Reich, Patterson et al. 2012, Raghavan, DeGiorgio et al. 2014) the Saqqaq individual, and Paleo-Eskimos in general,(Raghavan, DeGiorgio et al. 2014) represent a separate and relatively recent migration into America. Paleo-Eskimos demonstrate large proportions of Beringian (i.e. Chukotko-Kamchatkan and Eskimo-Aleut), Siberian, and South-East Asian ancestry. We also show that the Kets and closely associated Selkups belong to a group of modern populations closest to an ancient source of Siberian ancestry in Saqqaq. This group also includes, but is probably not restricted to, Uralic-speaking Nganasans, Yukaghirs (speaking an isolated language), and Tungusic-speaking Ulchi and Evens. Unlike the other populations of this group, the Kets, and, to a lesser degree the Selkups, have a high proportion of Mal'ta (ancient North Eurasian) ancestry.

As shown previously,(Reich, Patterson et al. 2012) Chipewyans, a modern Na-Dene-speaking population, have about 16% of Saqqaq ancestry. Thus, a gene flow, most likely post-dating the initial settlement of America 5,000-6,000 YBP(Rasmussen, Li et al. 2010) can be traced from the cluster of Siberian populations to Saqqaq, and from Saqqaq to Na-Dene. However, the genetic signal in contemporary Na-Dene-speaking ethnic groups is substantially diluted. The genetic proximity of Kets to the source of Siberian ancestry in Saqqaq correlates with the hypothesis that Na-Dene languages of North America are specifically related to Yeniseian languages of Siberia, now represented only by the Ket language.(Vajda 2011) However, this genetic link is indirect, and requires further study of population movement and language shifts in Siberia.

## Materials and Methods

*Sample Collection*

Saliva samples were collected and stored in the lysis buffer (50 mM Tris, 50 mM EDTA, 50 mM sucrose, 100 mM NaCl, 1% SDS, pH 8.0) according to the protocol of Quinque et al.(Quinque, Kittler et al. 2006) The buffer was divided into 3 mL aliquots in sterile 15 mL tubes. The following cities and villages along the Yenisei River were visited due to their accessibility either by boat or by helicopter (Suppl. Fig. 1.1): Dudinka (69.4°, 86.183°), Ust'-Avam (71.114°, 92.821°), Volochanka (70.976°, 94.542°), Potapovo (68.681°, 86.279°), Farkovo (65.720°, 86.976°), Turukhansk (administrative center of the Turukhanskiy district, 65.862°, 87.924°), Baklanikha (64.445°, 87.548°), Maduika (66.651°, 88.428°), Verkhneimbatsk (63.157°, 87.966°), Kellog (62.489°, 86.279°), Bakhta (68.841°, 96.144°), Bor (61.601°, 90.018°), Sulomai (61.613°, 91.180°). Volunteers rinsed their mouths with cold boiled water, and then collected up to 2 mL of saliva into a tube filled with the buffer. Samples were stored at environmental temperature (ranging from 4°C to 30°C) for up to two weeks, and then transported to a laboratory. DNA was isolated within one month after sample collection. Information about ethnicity, place of birth, and about first-, second-, and third-degree relatives was provided by the volunteers. All volunteers have signed informed consent forms, and the study was approved by the ethical committee of the Lomonosov Moscow State University (Russia) and supported by local administrations of the Taymyr and Turukhansk districts. In addition, the study was discussed with local committees of small Siberian nations for observance of their rights and traditions.

*DNA extraction*

DNA extraction protocol was adapted from the high-salt DNA extraction method(Quinque, Kittler et al. 2006). Fifteen microliters of proteinase K (20 mg/mL, Sigma) and 40 μL of 20% SDS were added to 1 mL of saliva in the buffer mixture, which was then incubated overnight at 53°C in a solid thermostat. After addition of 200 μL of 5M NaCl and incubation for 10 min on ice, the mixture was centrifuged for 10 min at 13,000 rpm in an Eppendorf 5415D centrifuge. The supernatant from each tube was transferred to a new tube, to which an equal amount of isopropanol was added. The tubes were then incubated for 10 min at room temperature and centrifuged for 15 min at 13,000 rpm. The supernatants were discarded, and the pellets washed once with 400 μL of 70% ethanol; then the pellets were dried and dissolved in 30-70 μL of sterile double-distilled water. The quantity of total DNA in samples was checked with Qubit® (Life Technologies, USA). DNA extracted from saliva represents a mix of human and bacterial DNA, and their ratio was checked by quantitative (q) PCR with 2 primer pairs (human: 1e1 5'-GTCCTCAGCGCTGCAGACTCCTGAC-3', BG1R 5'-CTTCCGCATCTCCTTCTCAG-3'; bacterial: 8F 5'-AGAGTTTGATCCTGGCTCAG-3', 519R 5'-GWATTACCGCGGKGCTG-3'). The PCR reaction mixture included: 13,2 μL of sterile water, 5 μL of qPCR master mix PK154S (Evrogen, Moscow, Russia), 0,4 μL of each primer and 1 μL of DNA sample. Amplification was performed with thermocycler StepOnePlus Applied Biosystems™ (Life Technologies, USA) with the standard program. Most DNA samples had low levels of bacterial contamination, and were used for further analysis: 22% of samples had the human/bacterial DNA ratio <1; 59% of samples had the ratio from 1 to 2, 19% of samples had the ratio >2.

*Genotyping*

GenoChip (the Genographic Project's genotyping array),(Elhaik, Greenspan et al. 2013) was used for genotyping 158 related and unrelated individuals of mixed and non-mixed ethnicity sampled in this study (see a list of sample ethnicity, gender, locations and geographic coordinates in Suppl. file S1) . The GenoChip includes ancestry-informative markers obtained for modern populations, the ancient Saqqaq genome, and two archaic hominins (Neanderthal and Denisovan), and was designed to identify all known Y-chromosome and mitochondrial haplogroups. The chip was carefully vetted to avoid inclusion of medically relevant markers, and SNP selection was performed with the goal of maximizing pairwise $F_{ST}$. The chip allows genotyping about 12,000 Y-chromosomal and approximately 3,300 mitochondrial SNPs, and over 130,000 autosomal and X-chromosomal SNPs. Genotyping was performed at the GenebyGene sequencing facility (TX, USA).

*Control of Sex Assignment*

In order to avoid mix-ups in sex assignment, we compared heterozygosity of X chromosome and missing rate among Y-chromosomal SNPs across the samples. All female samples had >62% Y-chromosomal SNPs missing and X chromosome heterozygosity >0.138, while male samples demonstrated values <1.2% and <0.007, respectively. Four wrong sex assignments were corrected based on these thresholds.

*Genome sequencing and genotype calling*

Genome sequencing has been performed for Ket individuals 884 (a male born in Baklanikha, mitochondrial haplogroup H, Y-chromosomal haplogroup Q1a2a1) and 891 (a female born in Surgutikha, mitochondrial haplogroup U5a1d). Prior to genome sequencing, we used NEBNext Microbiome DNA Enrichment Kit (New England Biolabs, USA) in order to enrich the samples for human DNA. This kit exploits the difference in methylation between eukaryotes and prokaryotes through selective binding of CpG-methylated (eukaryotic) DNA by the MBD2 protein. We took 500 ng of DNA from each sample and processed it according to manufacturer's instructions. Human DNA was eluted from magnetic beads using Proteinase K (Fermentas, Lithuania). Success of the enrichment was estimated using qPCR with primers for human RPL30 and bacterial 16S rRNA genes. The resulting human-enriched fraction was used for library construction using TruSeq DNA sample preparation kit (Illumina, USA). In parallel we made libraries from non-enriched DNA in order to assess whether enrichment leads to biases in sequence coverage. Libraries from enriched and non-enriched DNA were sequenced using the HiSeq2000 instrument (Illumina, USA) with read length 101+101 bp, two lanes for each library. As both enriched and non-enriched libraries produced similar coverage profiles and similar SNP counts in test runs of the *bcbio-nextgen* genotype calling pipeline (data not shown), their reads were pooled for subsequent analyses. Resulting read libraries for samples 884 and 891 had a median insert size of 215 bp and 343 bp, and coverage of 61x and 44x, respectively.

The *bcbio-nextgen* pipeline v. 0.7.9 (https://bcbio-nextgen.readthedocs.org/en/latest/) has been used with default settings for the whole read processing workflow: adapter trimming, quality filtering, read mapping on the reference genome hg19 with BWA, duplicate removal with Picard, local realignment, SNP calling, and recalibration with GATK v3.2-2, and annotation against dbSNP_138 with snpEff. Two alternative genotype calling modes have been tested in GATK: batch genotype calling for several samples, emitting only sites with at least one non-reference allele in at least one individual (GATK options --standard_min_confidence_threshold_for_calling 30.0 --standard_min_confidence_threshold_for_emitting 30.0, --emitRefConfidence at default); or calling genotypes for each sample separately, emitting all sites passing the coverage and quality filters (GATK options --standard_min_confidence_threshold_for_calling 30 --standard_min_confidence_threshold_for_emitting 30 --emitRefConfidence GVCF --variant_index_type LINEAR --variant_index_parameter 128000). The former approach maximized the output of dbSNP-annotated sites (7.36-7.62 million sites per individual vs. ~3.7 million sites for the latter approach), and therefore was used to generate calls subsequently merged with various SNP array and genomic datasets (Suppl. Table 1). To increase sensitivity of this approach, genotype calling has been performed with GATK HaplotypeCaller for six genomes in one run: Kets 884 and 891, two Yoruba, and two Kinh (Vietnamese) individuals downloaded from the 1000 Genome Project database(Genomes Project, Abecasis et al. 2012). We chose Yoruba samples NA19238 and NA19239, and Vietnamese samples HG01873 and HG02522 as they had read coverage similar to the Ket samples.

*Combined Datasets*

Combined datasets generated in this study and their properties are listed in Suppl. Table 1. In all cases, datasets were designed to maximize population coverage in Russia (Siberia in particular), in Central Asia and in the Americas, while keeping only few reference populations in the Middle East, in South and South-East Asia, Africa, Australia, and Oceania. Third-degree and closer relatives detected through questionnaires and pedigree analysis and individuals of mixed ethnicity were excluded from the Enets, Ket, Nganasan, and Selkup population samples. Overall, 88 of 158 individuals remained (see supposed percentage of relatedness for each sample pair and a list of selected samples in Suppl. file S1 and hierarchical clustering of all samples based on genetic distance in Suppl. Fig. 4.1). All datasets underwent filtering using PLINK(Purcell, Neale et al. 2007) v. 1.9. Maximum missing rate per SNP thresholds of 0.03 or 0.05 were used (Suppl. Table 1), except for the full-genome dataset 'Ket genomes + reference genomes', for which a more relaxed threshold of 0.1 was used to accommodate the Mari individual with low coverage(Raghavan, Skoglund et al. 2014) and to keep the high number of SNPs at the same time. Including the Mari individual was considered important since various analyses on other datasets grouped Mari (and Uralic-speaking populations in general) and Kets together. Linkage disequilibrium (LD) filtering was applied to all datasets except for the GenoChip-based ones since SNPs included into the GenoChip array underwent LD filtering with the $r^2$ threshold of 0.4 as described in Elhaik et al.(Elhaik, Greenspan et al. 2013) For the other datasets the following LD filtering settings were used: window size of 50 SNPs, window step of 5 SNPs, $r^2$ threshold 0.5 (PLINK v. 1.9 option '--indep-pairwise 50 5 0.5'). Ancient genome data within the HumanOrigins dataset were provided as artificially haploid: one allele was selected randomly at each diploid site, since confident diploid calls were not possible for low-coverage ancient genomes (see further description of the approach in Lazaridis et al.(Lazaridis, Patterson et al. 2014)). We applied the same procedure to ancient genomes included into dataset 'Ket genomes + reference genomes'. Other details pertaining to individual datasets are listed below, and their population composition is shown in Suppl. Table 2. We prepared five additional datasets based on two Ket genomes sequenced in this study and the Mal'ta and Saqqaq ancient genomes: three datasets of various population composition including the HumanOrigins SNP array data (69K, 195K, and 196K SNPs), a genome-based dataset (347.5K SNPs), and its version with transition, i.e. CT and AG, polymorphisms removed (185K SNPs) (Suppl. Tables 1 and 2). Taking into account admixture coefficients for the two sequenced Ket individuals (Ket891 and Ket884, Fig. 1A), we selected Ket891 as an individual with lower values of the North European and Siberian admixture components (in the K=19 dimensional space). In addition, Ket891 was identified as non-admixed by reAdmix analyses (Suppl. Table 3).

**GenoChip** (alternative name 'GenoChip-only'). The dataset was taken from Elhaik et al.(Elhaik, Tatarinova et al. 2014) and merged with genotype calls obtained with GenoChip in this study. After filtering, the dataset contained 732 individuals from 47 populations and 116,916 SNPs, and had low missing SNP rates (maximum missing rate per individual 0.033).

**GenoChip + Illumina arrays** (alternative name 'GenoChip-based'). The dataset was constructed by merging selected populations from dataset GenoChip, genotyping data obtained with GenoChip in this study, and SNP array data (various Illumina models) from the following sources: Behar et al.(Behar, Yunusbayev et al. 2010); Cardona et al.(Cardona, Pagani et al. 2014); Fedorova et al.(Fedorova, Reidla et al. 2013); Li et al.(Li, Absher et al. 2008); Kidd et al(Kidd, Friedlaender et al. 2011); Raghavan et al.(Raghavan, Skoglund et al. 2014); Rasmussen et al.(Rasmussen, Li et al. 2010); Reich et al.(Reich, Patterson et al. 2012); Silva-Zolezzi et al.(Silva-Zolezzi, Hidalgo-Miranda et al. 2009); Surakka et al.(Surakka, Kristiansson et al. 2010); Yunusbayev et al.(Yunusbayev, Metspalu et al. 2012) Three ancient genomes, La Braña(Olalde, Allentoft et al. 2014), Saqqaq(Rasmussen, Li et al. 2010), and Clovis(Rasmussen, Anzick et al. 2014), were also added (genotypes in VCF format were obtained from the respective publications). Only SNPs included into the GenoChip array were used and further filtered as described above. Maximum missing rate per individual was 50%, but three ancient genomes (Clovis, Saqqaq and La Braña) were exempt. After filtering, the dataset contained 1,624 individuals from 90 populations and 32,189 SNPs.

**Ket genomes + Illumina arrays**. In order to include populations relevant for our analyses, e.g. Burusho, Khanty, and Nenets, omitted from the previous dataset due to very low marker overlaps, full-genome SNP calls for two Ket individuals (see above) were merged with SNP array data (various Illumina models) from the following sources: HapMap3(International HapMap, Altshuler et al. 2010); Behar et al.(Behar, Yunusbayev et al. 2010, Behar, Metspalu et al. 2013); Cardona et al.(Cardona, Pagani et al. 2014);

Fedorova et al.(Fedorova, Reidla et al. 2013); Li et al.(Li, Absher et al. 2008); Raghavan et al.(Raghavan, Skoglund et al. 2014); Rasmussen et al.(Rasmussen, Li et al. 2010); Reich et al.(Reich, Patterson et al. 2012); Silva-Zolezzi et al.(Silva-Zolezzi, Hidalgo-Miranda et al. 2009); Yunusbayev et al.(Yunusbayev, Metspalu et al. 2012) The filtered dataset contained modern individuals only: 2,549 individuals from 105 populations and 103,495 SNPs, and had low missing SNP rates (maximum missing rate per individual 0.04).

**Ket genomes + HumanOrigins array** (alternative name 'HumanOrigins-based'). For analyzing relevant ancient genomes alongside the Ket genomes in context of multiple modern populations genotyped with the HumanOrigins Affymetrix SNP array(Patterson, Moorjani et al. 2012, Lazaridis, Patterson et al. 2014), the full dataset of Lazaridis et al.(Lazaridis, Patterson et al. 2014) was merged with Ket genome data and filtered (LD filtering with the $r^2$ threshold of 0.5, maximum per SNP missing rate 0.05). The resulting set contained 217 populations/genomes. In order to make the dataset more manageable computationally and more focused, 78 populations/genomes were removed, including: gorilla, orangutan, macaque, and marmoset genomes; low-coverage (<1x) ancient genomes of anatomically modern humans [Afontova Gora-2/AG2(Raghavan, Skoglund et al. 2014), a west European hunter-gatherer and a farmer(Skoglund, Malmstrom et al. 2012), Motala hunter-gatherers(Lazaridis, Patterson et al. 2014)]; the Iceman genome (a west European ancient farmer(Keller, Graefen et al. 2012)); human reference genome hg19. The following ancient genomes were included: Neanderthal and Denisovan genomes; La Braña 1(Olalde, Allentoft et al. 2014), Loschbour(Lazaridis, Patterson et al. 2014), Mal'ta/MA1(Raghavan, Skoglund et al. 2014), and Motala12(Lazaridis, Patterson et al. 2014) (Eurasian hunter-gatherers); Stuttgart (a west European farmer of the LBK archaeological culture(Lazaridis, Patterson et al. 2014)) and Saqqaq(Rasmussen, Li et al. 2010). The final dataset contained 1,786 individuals from 139 populations and 195,918 SNPs.

**Ket genomes + HumanOrigins array, reduced**. For performing TreeMix analysis on the HumanOrigins-based dataset, the original dataset was reduced to 39 most relevant populations prior to filtering. Population composition was designed to maximize overlap with that of the full-genome dataset (see below), also used for TreeMix analysis, and to include as many Siberian populations as possible. Then filtering was applied: LD filtering with the $r^2$ threshold of 0.5, maximum per SNP missing rate of 0.05. The final dataset contained 527 individuals from 39 populations and 194,750 SNPs.

**Ket genomes + HumanOrigins array + Verdu et al. 2014**. The final version of the previous dataset 'Ket genomes + HumanOrigins array' was merged with Illumina 610-Quad SNP array genotyping data for six North American native populations (Haida, Nisga'a, Splatsin, Stswecem'c, Tlingit, Tsimshian(Verdu, Pemberton et al. 2014)) and filtered (Suppl. Table 1). The major reason for constructing this dataset was the inclusion of Tlingit and Haida, Na-Dene-speaking populations not present in the other datasets. The final dataset contained 1,867 individuals from 145 populations and 68,625 SNPs.

**Ket genomes + reference genomes** (alternative name 'full-genome'). The following seven ancient genomes were included into the dataset: Clovis(Rasmussen, Anzick et al. 2014), Late Dorset(Raghavan, DeGiorgio et al. 2014) and Saqqaq(Rasmussen, Li et al. 2010) Paleo-Eskimos, Mal'ta (an ANE representative(Raghavan, Skoglund et al. 2014)), Motala12 and Loschbour (WHG(Lazaridis, Patterson et al. 2014)), Stuttgart (EEF(Lazaridis, Patterson et al. 2014)). To ensure dataset uniformity, genotype calling for these ancient genomes was performed *de novo* in a batch run, instead of using published genotypes generated with different genotype calling protocols. Ancient DNA reads mapped on the reference genome hg19 (provided by their respective authors) were used for genotype calling with the ANGSD software v. 0.800(Korneliussen, Albrechtsen et al. 2014) with the following settings: SAMtools calling mode (option -GL 1); genotype likelihood output (option -doGlf 2); major allele specified according to the reference genome (-doMajorMinor 4); allele frequency obtained based on the genotype likelihoods (-doMaf 1); SNP *p*-value $10^{-6}$. The resulting genotype likelihood files were transformed into genotypes in the VCF format using BEAGLE utilities gprobs2beagle and beagle2vcf, with a minimum genotype likelihood cut-off of 0.6. Subsequently one allele was selected randomly at each diploid site, since confident diploid calls were not possible for low-coverage ancient genomes. Genotype data for ancient genomes were merged with the following modern samples using PLINK v. 1.9 (and subsequently filtered as described in Suppl. Table 1):(i) 7.36-7.62 million GATK genotype calls for two Kets, two Yoruba, and two Vietnamese individuals (sites with a non-reference allele in at least one individual, see above); and (ii) genotypes at both homozygous reference and non-reference sites for: 1 Aleutian, 2 Athabaskans, 2 Greenlanders, 2

Nivkhs(Raghavan, DeGiorgio et al. 2014); 1 Avar, 1 Indian, 1 Mari, 1 Tajik(Raghavan, Skoglund et al. 2014); 1 Australian aboriginal,(Rasmussen, Guo et al. 2011) 1 Karitiana, 1 Mayan(Rasmussen, Anzick et al. 2014); Simons Genome Diversity Project panels A(Meyer, Kircher et al. 2012) and B(Prufer, Racimo et al. 2014) containing in total 25 genomes of 13 populations. The final dataset contained 52 individuals from 31 populations and 347,466 SNPs.

**Ket genomes + reference genomes without transversions**. In order to mitigate the effect of ancient DNA deamination and the resulting C to T substitutions(Axelsson, Willerslev et al. 2008), we constructed another version of the full-genome dataset, with all CT and AG SNPs excluded prior to the LD filtering step. The final dataset contained 52 individuals from 31 populations and 185,382 SNPs.

*PCA*

The principal component analysis (PCA) was carried out in the smartpca program included in the EIGENSOFT package(Patterson, Price et al. 2006). We calculated 10 eigenvectors and ran the analysis without removing outliers.

*ADMIXTURE analysis*

The ADMIXTURE software implements a model-based Bayesian approach that uses block-relaxation algorithm in order to compute a matrix of ancestral population fractions in each individual (Q) and infer allele frequencies for each ancestral population (P)(Alexander, Novembre et al. 2009). A given dataset is usually modeled using various numbers of ancestral populations (K). Here we used the unsupervised admixture approach, in which allele frequencies for non-admixed ancestral populations are unknown and are computed during the analysis. For each K from 2 to 25, 100 analysis iterations were generated with different random seeds. The best run was chosen according to the highest log likelihood. For each run 10-fold cross-validation (CV) was computed.

*TreeMix analysis*

Maximum likelihood tree construction and admixture modelling was performed with the TreeMix v. 1.12 software(Pickrell and Pritchard 2012) on the full-genome dataset of 31 populations and 52 individuals: on its original version or the version with CT and AG SNPs excluded. Initially, SNP window length (k) was optimized by testing various values (k=1, 5, 10, 20, 50, or 100) with 10 different random seeds, and the k setting producing the highest percentage of variance explained by the model was selected. Final runs were performed with the following settings: SNP window length, 5; number of migration edges from 0 to 10; trees rooted with the San population; global tree rearrangements used (option '-global'); no sample size correction (option '-noss'); 100 iterations, selecting a tree with the highest likelihood (and with the highest explained variance percentage among trees with identical likelihoods). No pre-defined migration events were incorporated. Residuals from the fit of the model to the data were plotted and percentage of explained variance was calculated with scripts supplied in the TreeMix package. One hundred bootstrap replicates re-sampling blocks of 5 SNPs were calculated and respective trees were constructed for 8, 9, and 10 migration edges using TreeMix. Bootstrap support values for nodes were mapped on the original trees using RAxML. Bootstrap values for migration edges were interpreted as described in Suppl. Information Section 9 and in the respective figure legends (Fig. 2).

*$f_3$, $f_4$, and D statistics*

We used three and four population tests ($f_3$ and $f_4$) developed by Patterson et al.(Patterson, Moorjani et al. 2012) and implemented in programs *threepop* and *fourpop* of the TreeMix(Pickrell and Pritchard 2012) package. The source code in $C^{++}$ was modified to enable multithreading and computing the statistic for all population combinations with a given population or population pair. SNP windows used for computing standard errors of $f_3$ and $f_4$ statistics are shown for each dataset in Suppl. Table 1. Statistic $f_3$(O; A, $X_1$)(Patterson, Moorjani et al. 2012) measures relative amount of genetic drift shared between the test population A and a reference population $X_1$, given an outgroup population O distant from both A and $X_1$. Outgroup $f_3$ statistic is always positive, and its values can be interpreted only in the context of the reference dataset X. Statistic $f_4$(X, O; A, B)(Patterson, Moorjani et al. 2012) tests whether A and B are equidistant from X, given a sufficiently distant outgroup O: in that case the statistic is close to zero. Otherwise, the statistic shows whether X is more closely related to A or to B, hence $f_4$ may be positive or negative. The statistical significance of $f_4$ values is typically assessed using a Z-score: an $f_4$ value divided by its standard deviation. A threshold Z-score of 1.96 (rounded to 2 in this paper) corresponds to a *p*-value of 0.05.

To estimate the Neanderthal gene flow influence we performed D-statistic analysis as described in Green et al.(Green, Krause et al. 2010) Reads for two Yoruba and two Kinh (Vietnamese) individuals were downloaded from the 1000 Genome Project database.(Genomes Project, Abecasis et al. 2012) We chose Yoruba samples NA19238 and NA19239, and Kinh Vietnamese samples HG01873 and HG02522 as they had read coverage similar to the Ket samples, and were not genetically related to each other. Ket, Yoruba, and Vietnamese reads were used for calling SNPs with GATK HaplotypeCaller, emitting both reference and non-reference sites, about 1 billion sites per individual. This procedure ensured that genotype calls for each individual were made in exactly the same way. Altai Neanderthal and chimpanzee genotypes were processed as described in Khrameeva et al.(Khrameeva, Bozek et al. 2014) Coordinates of the chimpanzee genome were mapped to the human genome hg19 using UCSC liftOver tool.(Rosenbloom, Armstrong et al. 2015)

In further analysis, we considered only homozygous sites different between the chimpanzee (A) and Neanderthal (B) genomes. Then we matched a randomly selected modern human allele to these sites. All sites where a Ket allele matched a Neanderthal allele and a Yoruba allele matched a chimpanzee allele were counted and referred to as #ABBA (termed Neanderthal-like sites). All sites where a Ket allele matched a chimpanzee allele and a Yoruba allele matched a Neanderthal allele were counted and referred to as #BABA. *D*-statistic = (#ABBA – #BABA)/(#ABBA + #BABA) was calculated and averaged for all possible pairs of Yoruba and Ket samples. As a control, the same analysis was repeated for Vietnamese genotypes instead of Ket genotypes.

Ket and Vietnamese sites used in the *D*-statistic analysis were assigned to human genes according to coordinates of the longest transcript retrieved from UCSC Genome Browser(Rosenbloom, Armstrong et al. 2015) plus 1,000 nucleotides upstream to include potential regulatory regions. The gene set enrichment analysis (GSEA) algorithm(Subramanian, Tamayo et al. 2005) ranked genes according to difference between #ABBA and #BABA, while four pairs of samples were treated as replicates. We used the MSigDB collection of 825 gene ontology (GO) biological processes (c5.bp.v3.0.symbols.gmt)(Subramanian, Tamayo et al. 2005) to assign genes to functional groups. GO terms with less than 15 or more than 500 genes per term were excluded. The mean and median false discovery rates (the mean FDR and median FDR) were used to estimate the significance of Neanderthal sites enrichment in the functional groups. In GSEA, the mean FDR was obtained by using the mean of the estimated number of false positives in each of 3000 permutations of the sample labels, while the median FDR was calculated as the median of the estimated number of false positives in the same permutations.

*Clustering*

Within the Ket population, we have found a number of subpopulations using a combination of KMEANS clustering and Kullback-Leibler distance approach(Sahu and Cheng 2003). We used the KMEANS clustering routine in *R*. Let N be the number of individuals. We ran the KMEANS clustering for *k* ranging from the N to two, using the matrix of admixture proportions as input (the matrix was calculated with ADMIXTURE(Alexander, Novembre et al. 2009) for the dataset GenoChip). At each iteration, we calculated the ratio of the sum of squares between groups and the total sum of squares. If this ratio was >0.9, then we accepted the *k*-component model. Since KMEANS clustering cannot be implemented for *k*=1, to decide between two clusters or a possible single cluster, we also calculated Kullback-Leibler distance (KLD) between the *k*=2 and *k*=1 models. If the KLD <0.1 and the ratio of the sum of squares between groups and the total sum of squares for two-component model was above 0.9, then the *k*=1 model was selected because, in such cases, there were no subgroups in the population.

*GPS*

An admixture-based Geographic Population Structure (GPS) method(Elhaik, Tatarinova et al. 2014) was used for predicting the provenance of all genotyped individuals (including relatives). GPS finds a global position where the individuals with the genotype closest to the tested one live. GPS is not suitable to analyzed recently admixed individuals. GPS calculated the Euclidean distance between the sample's admixture proportions and the reference dataset. The matrix of admixture proportions was calculated with ADMIXTURE(Alexander, Novembre et al. 2009) for dataset GenoChip. The shortest distance, representing the test sample's deviation from its nearest reference population, was subsequently converted into geographical distance using the linear relationship observed between genetic and geographic distances. The final position of the sample on the map was calculated by a linear combination of vectors, with the origin at the geographic center of the best matching population weighted by the distances to 10 nearest reference populations and further scaled to fit on a circle with a radius proportional to the geographical distance.

*reAdmix*

reAdmix(Kozlov, Chebotarov et al. 2015) estimates individual mixture in terms of present-day populations and operates in unconditional and conditional modes. reAdmix models ancestry as a weighted sum of present-day populations (e.g. 50% British, 25% Russian, 25% Han Chinese) based on the individual's admixture components. In conditional mode, the user may specify one or more known ancestral populations, and in unconditional mode, no such information is provided. We used reAdmix for analysis of the Ket, Selkup, Nganasan, and Enets samples in unconditional mode, and the matrix of admixture proportions was calculated with ADMIXTURE(Alexander, Novembre et al. 2009) for dataset GenoChip.

*Prediction of mitochondrial and Y-chromosome haplogroups*

Mitochondrial genome SNPs (approximately 3,300) were genotyped with the GenoChip array in 158 individuals. SNP loci heterozygous in more than 15 samples or those with missing data in more than 15 samples were removed completely, and remaining heterozygous genotypes were filtered out in particular individuals. Mitochondrial DNA haplogroups were predicted using the MitoTool software (http://www.mitotool.org/).

SNPs typed on Y chromosome with the GenoChip array were checked and low-quality SNPs with genotyping rate <95% were removed for all 53 male individuals genotyped with GenoChip in this study. One sample (sample ID GRC14460103) was removed due to poor genotyping rate (18.7% missing markers on Y chromosome). After this quality control step, 11,883 high-quality Y-chromosomal SNPs remained for the downstream analysis. Genotyping data were transformed into a list of mutations and haplogroup prediction was performed using the Y-SNP Subclade Predictor online tool at MorleyDNA.com (http://ytree.morleydna.com/).

## Acknowledgements

We are grateful to all sample donors, and to local community members for their help in sample collection. We thank Eske Willerslev, Simon Rasmussen, Maanasa Raghavan, Iñigo Olalde, Andrés Ruiz-Linares, David Reich, Noah Rosenberg and Ripan Malhi for sharing genotyping and sequencing data. We would like to thank National Genographic and Family Tree DNA for genotyping our samples, Shi Yan (Fudan University, Shanghai, China) and Horolma Pamjav (Institute of Forensic Medicine, Budapest, Hungary) for their help with compiling Y-chromosome haplogroup frequency tables. Special thanks go to Alexey S. Kondrashov for putting our team together. P.F., P.C., and A.Z. were supported by the Moravian-Silesian region projects MSK2013-DT1, MSK2013-DT2, and MSK2014-DT1, and by the Institution Development Program of the University of Ostrava. T.V.T. was supported by grants from The National Institute for General Medical Studies (GM068968), the Eunice Kennedy Shriver National Institute of Child Health and Human Development (HD070996), and National Science Foundation Division of Evolutionary Biology (1456634). M.D.L. and M.S.G. were supported by the Russian Science Foundation: project nos. 14-50-00150 and 14-24-00155, respectively.

## Author contributions

P.F. and T.V.T. designed the study, took part in sample collection, performed data analyses, and wrote the paper. I.V.T., O.P.K. and T.N. were responsible for sample collection and manipulation, M.D.L. was responsible for genome sequencing, E.S.G. for software re-design, and A.Z., E.E.K., M.S.G., M.T., O.F., P.C., P.T., V.V.S. performed data analyses. G.S. contributed the linguistics section of the paper. Y.V.N. helped prepare the final version of the manuscript. All authors took part in interpretation and discussion of the results.

## Competing financial interests

The authors declare no competing financial interests.

## Figure legends

**Fig. 1. A.** Admixture coefficients plotted for dataset 'GenoChip + Illumina arrays'. Abbreviated names of admixture components are shown on the left as follows: SAM, South American; NAM, North American; ESK, Eskimo (Beringian); SEA, South-East Asian; SIB, Siberian; NEU, North European; ME, Middle Eastern; CAU, Caucasian; SAS, South Asian; OCE, Oceanian; AFR, African. The Ket-Uralic ('Ket') admixture component appears at $K \geq 11$, and admixture coefficients are plotted for K=10, 11, and 19. Although K=20 demonstrates the lowest average cross-validation error, the Ket-Uralic component splits in two at this K value, therefore K=19 was chosen for the final analysis. Only populations containing at least one individual with >5% of the Ket-Uralic component at K=19 are plotted, and individuals are sorted according to values of the Ket-Uralic component. Admixture coefficients for the Saqqaq ancient genome and for two Ket individuals sequenced in this study are shown separately on the right and on the left, respectively. **B.** Average cross-validation (CV) error graph with standard deviations plotted. Ten-fold cross-validation was performed. The graph has a minimum at K=20. **C.** Color-coded values of the Ket-Uralic admixture component at K=19 plotted on the world map. Maximum values in each population are taken, and only values >5% are plotted. Top five values of the component are shown in the bottom left corner, and the value for Saqqaq is shown on the map.

**Fig. 2. A.** A maximum likelihood tree with 17 migration edges computed on the reduced HumanOrigins-based dataset (Suppl. Table 1) with TreeMix (a tree with the highest likelihood was selected among 100 iterations). For clarity, only seven most relevant migration edges are visualized. Edge weight values are shown in the table, the drift parameter is shown on the x-axis, and bootstrap support values for tree nodes are indicated. **B.** Residuals from the fit of the model to the data visualized. 99.25% variance is explained by the tree. **C.** A maximum likelihood tree with 10 migration edges computed on the full-genome dataset (a tree with the highest likelihood was selected among 100 iterations). Edge weight and bootstrap support values are shown in the table, the drift parameter is shown on the x-axis, and bootstrap support values for tree nodes are indicated. Migration edges are numbered according to their order of appearance in the sequence of trees from *m*=0 to *m*=10. Notes to the figure: *As migration edges and tree topology are inter-dependent in bootstrapped trees, bootstrap support for the edges in the original tree was calculated by summing up support for closely similar edges in bootstrapped trees. Below these edge groups are listed for edges #1-10: 1/ Greenlander Inuit ⬄ Saqqaq and/or Late Dorset clade; 2/ (Australian, Papuan) clade ⬄ Kinh, (Dai, Kinh), (Han, Dai, Kinh) clades; 3/ Mari ⬄ Saqqaq and/or Late Dorset clade; 4/ Aleut ⬄ any clade composed only of European populations (considering Mal'ta a member of the European clade); 5/ Sardinian and/or Stuttgart clade ⬄ any African clade or a basal non-African clade; 6/ Loschbour ⬄ Mayan; 7/ Ket ⬄ any clade composed only of American and Beringian populations (excluding Saqqaq and Late Dorset); 8/ Ket ⬄ Saqqaq and/or Late Dorset clade; 9/ Ket ⬄ Motala12; 10/ Mari ⬄ Ket. ** Ket ⬄ Mal'ta migration edge appeared in 27 bootstrap replicates of 100. **D.** Residuals from the fit of the model to the data visualized. 98.79% variance is explained by the tree.

**Fig. 3. A.** PC3 vs. PC4 plot for the dataset 'Ket genomes+HumanOrigins array+Verdu et al. 2014'. African populations are not shown. Populations are color-coded by geographic region or language affiliation (in the case of Siberian and Central Asian populations), and most relevant populations are differentiated by marker shapes. Ancient genomes are shown in black. For the corresponding PC1 vs. PC2 plot see Suppl. Fig. 6.11. **B.** PC3 vs. PC4 plot, zoom on the Ket individuals.

**Fig. 4.** Statistics $f_3$(Yoruba; Mal'ta, X) computed on the full-genome dataset with individual Ket884 excluded. See the corresponding result for the dataset with transitions excluded in Suppl. Fig. 7.42. **A.** Color-coded $f_3$ values plotted on the world map. Top five values are shown in the bottom left corner. **B.** $f_3$ values (green circles) sorted in descending order with their standard errors shown by vertical lines.

**Fig. 5.** Statistics $f_4$(Mal'ta, Yoruba; Y, X) **(A)**, $f_4$(Ket884+891, Yoruba; Y, X) **(B)**, and $f_4$(Ket891, Yoruba; Y, X) **(C)** computed on the full-genome dataset with African, Australian and Papuan populations excluded. See the corresponding results for the dataset without transitions in Suppl. Figs. 8.35 and 8.37, respectively. A matrix of color-coded Z-scores is shown. Z-score equals the number of standard errors by which the statistic differs from zero. Populations are sorted in alphabetical order. Rows show Z-scores for $f_4$(Mal'ta, Yoruba; row, column) or $f_4$(Ket, Yoruba; row, column), *vice versa* for columns.

**Fig. 6.** Statistics $f_3$(Yoruba; Haida, X) computed on dataset 'Ket genomes+HumanOrigins array+Verdu et al. 2014' with individual Ket884 excluded. **(A)**. Color-coded $f_3$ values plotted on the world map. Top five values are shown in the bottom left corner. **(B)** Top $f_3$ values (green circles) sorted in descending order with their standard errors shown by vertical lines. Populations of America/Chukotka/Kamchatka and Eurasia are underlined by solid red and blue lines, respectively. **(C)** All $f_3$ values (green circles) sorted in descending order with their standard errors shown by vertical lines.

## Figures

Fig. 1, A

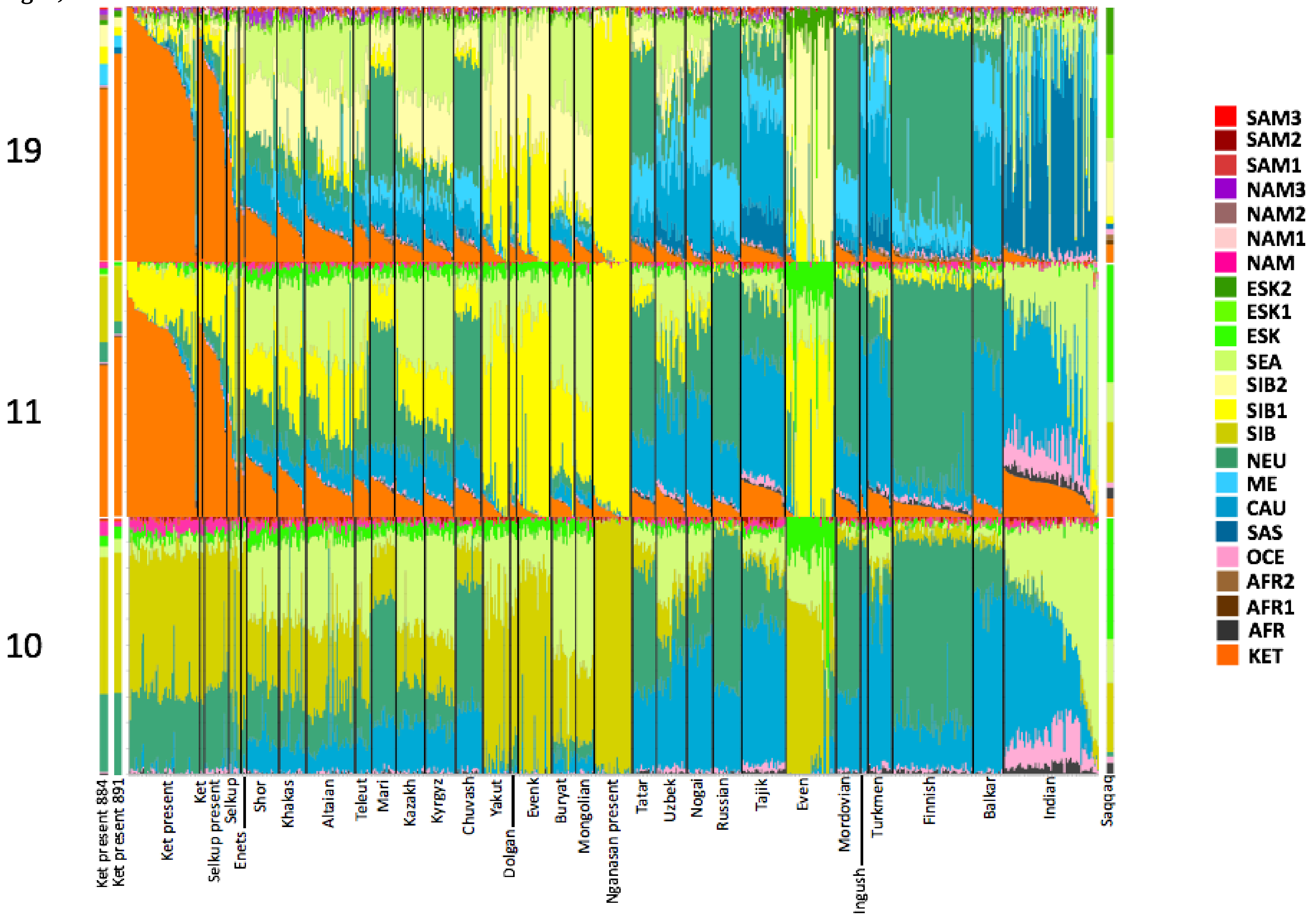

**Fig. 1, B**

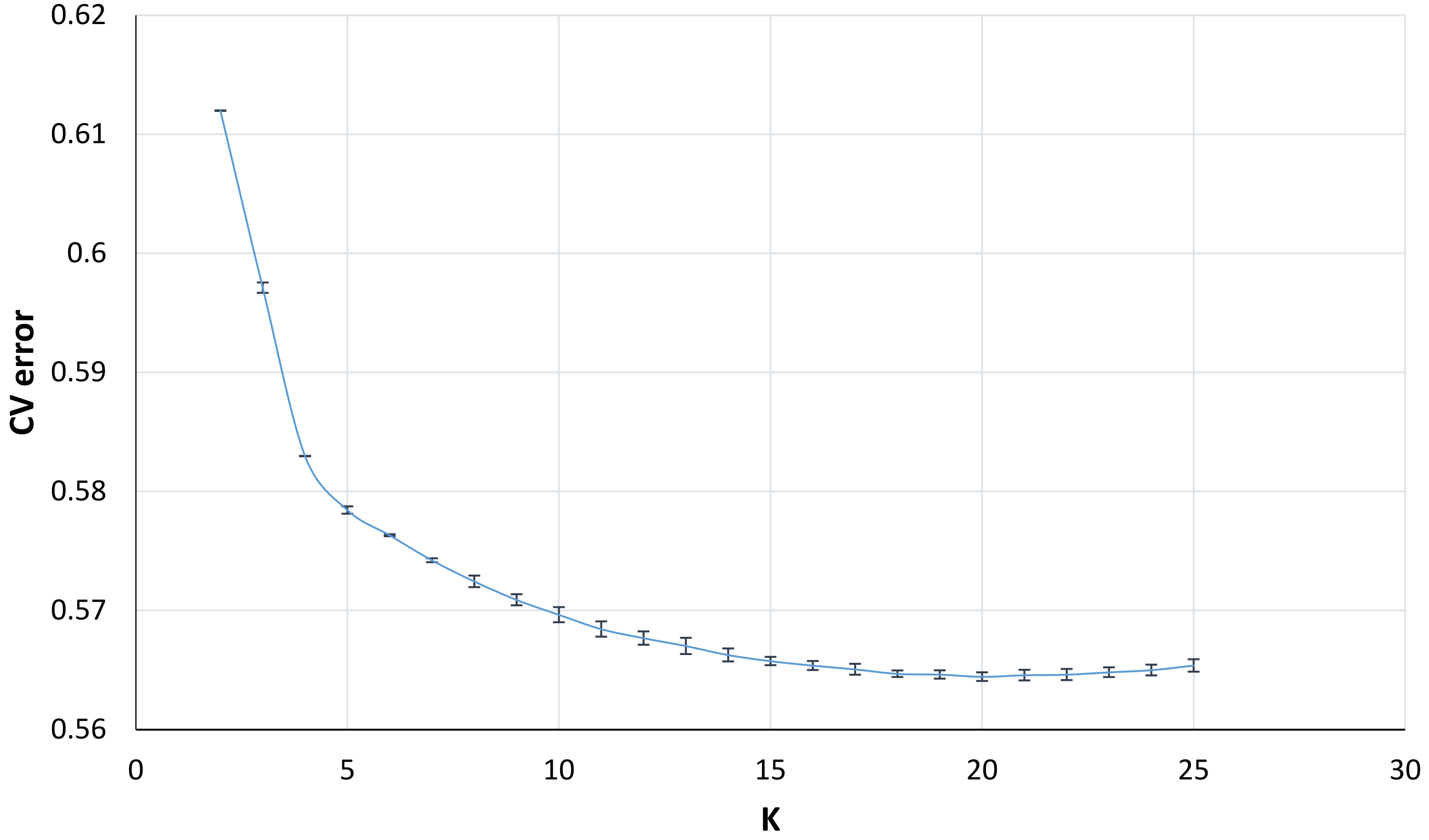

**Fig. 1, C**

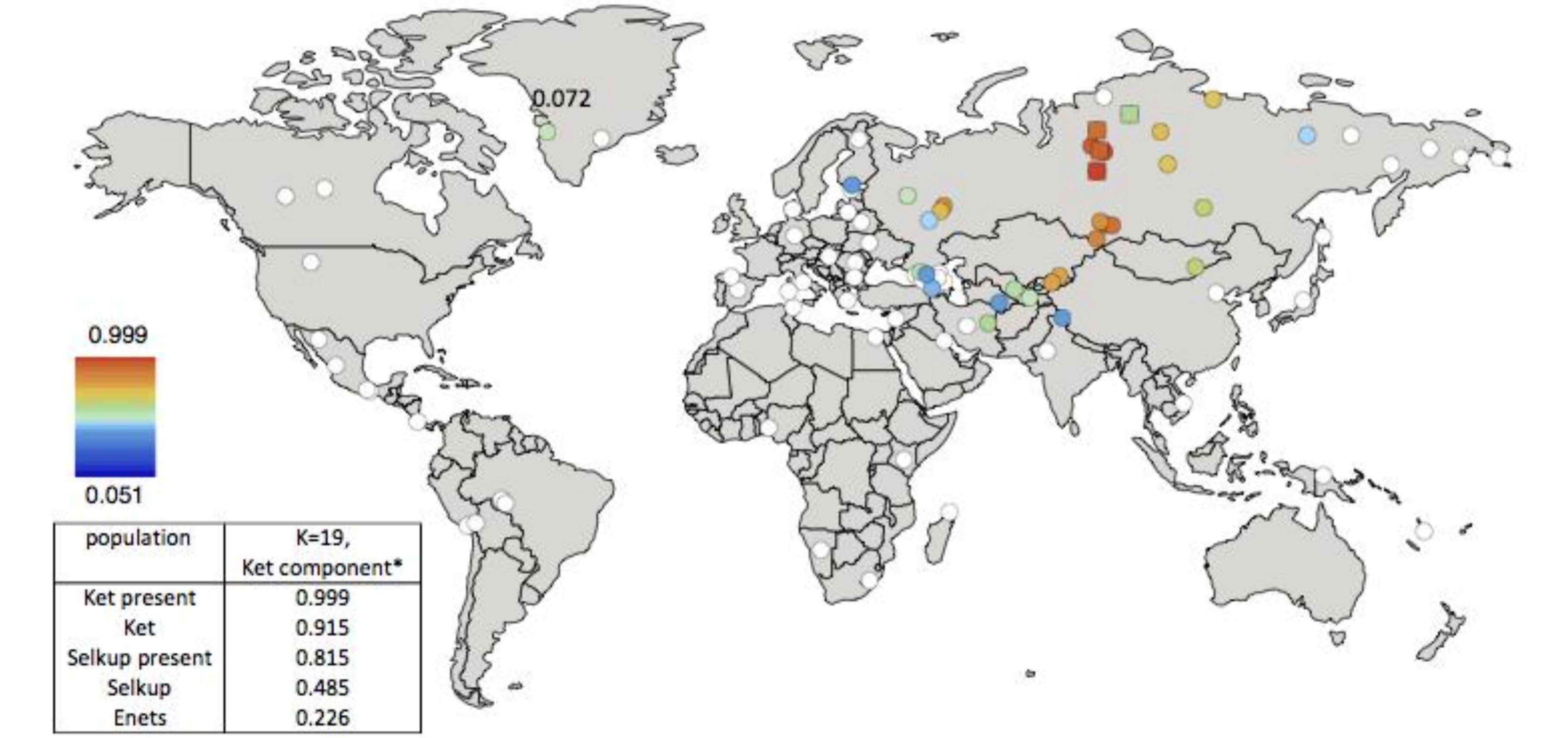

| population | K=19, Ket component* |
|---|---|
| Ket present | 0.999 |
| Ket | 0.915 |
| Selkup present | 0.815 |
| Selkup | 0.485 |
| Enets | 0.226 |

**Fig. 2, A.**

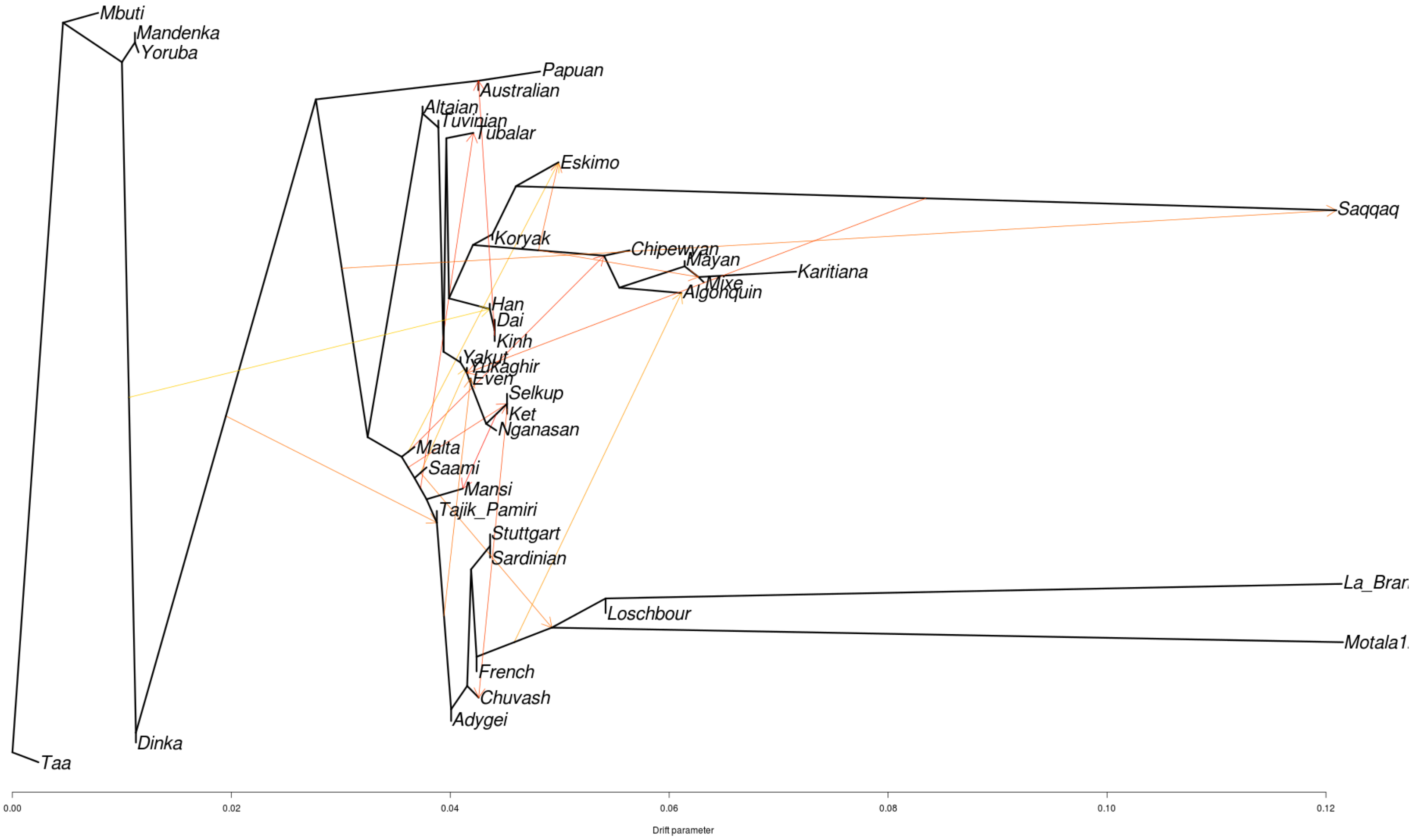

**Fig. 2, B.**

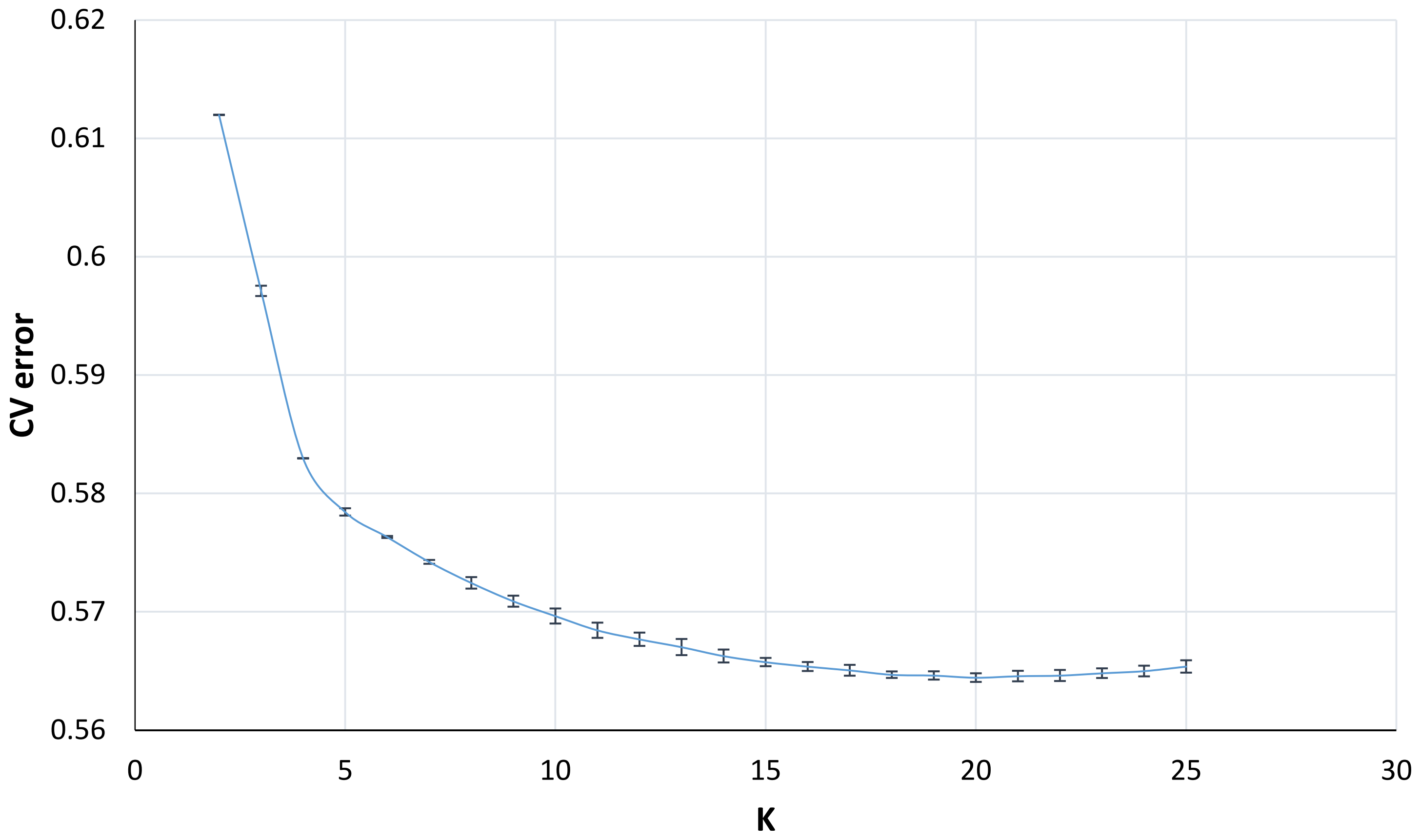

**Fig. 2, C.**

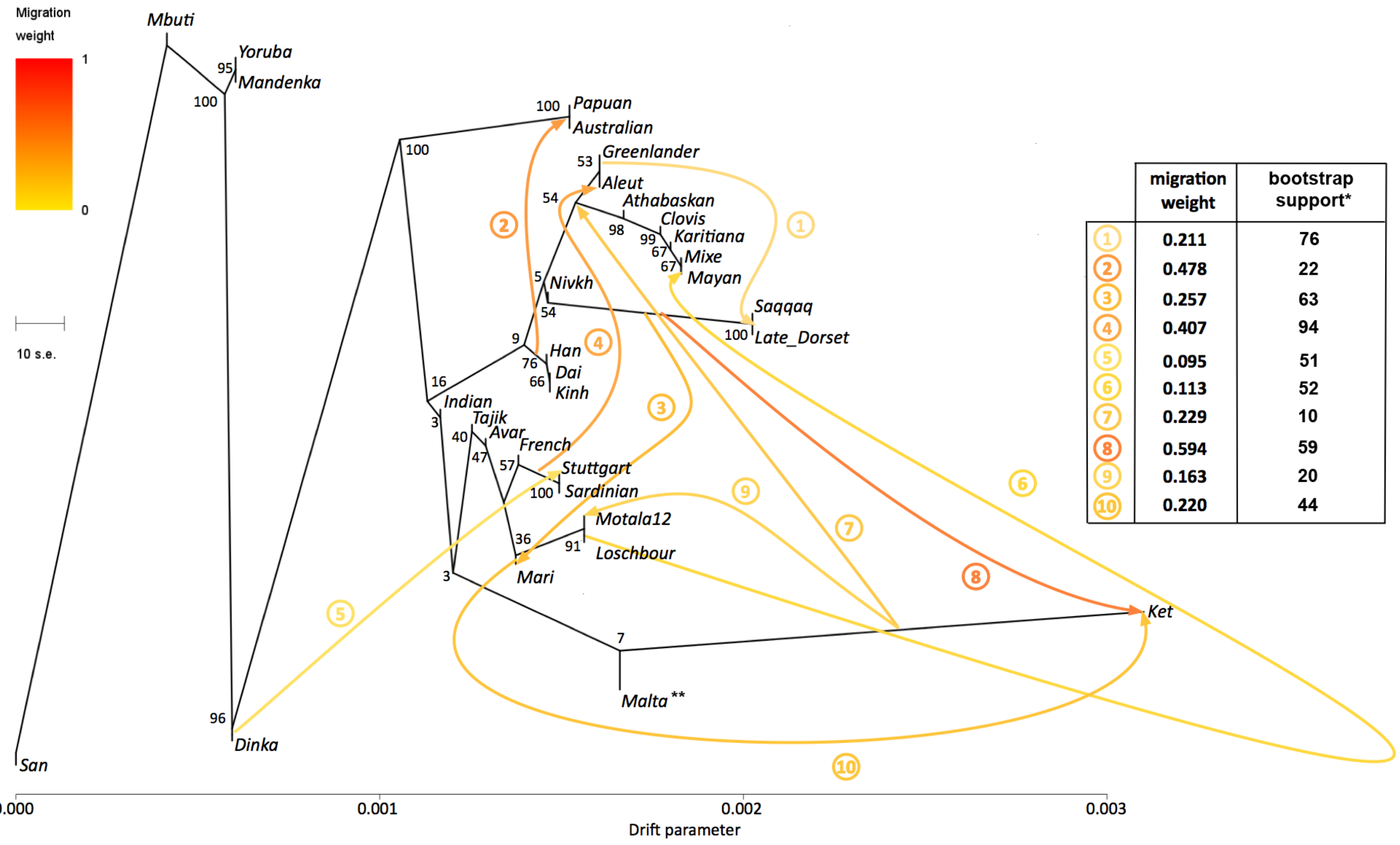

| | migration weight | bootstrap support* |
|---|---|---|
| 1 | 0.211 | 76 |
| 2 | 0.478 | 22 |
| 3 | 0.257 | 63 |
| 4 | 0.407 | 94 |
| 5 | 0.095 | 51 |
| 6 | 0.113 | 52 |
| 7 | 0.229 | 10 |
| 8 | 0.594 | 59 |
| 9 | 0.163 | 20 |
| 10 | 0.220 | 44 |

**Fig. 2, D.**

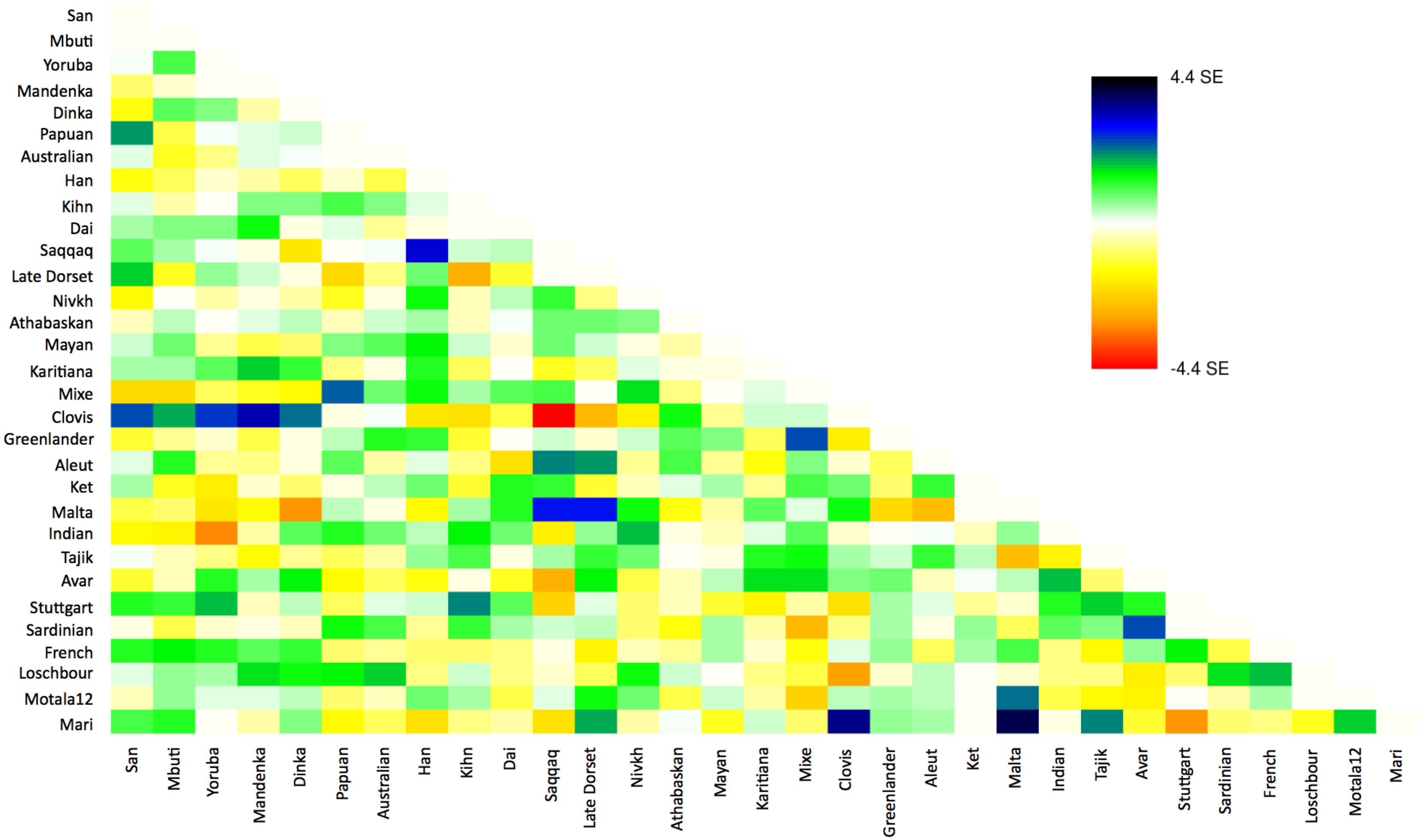

**Fig. 3, A.**

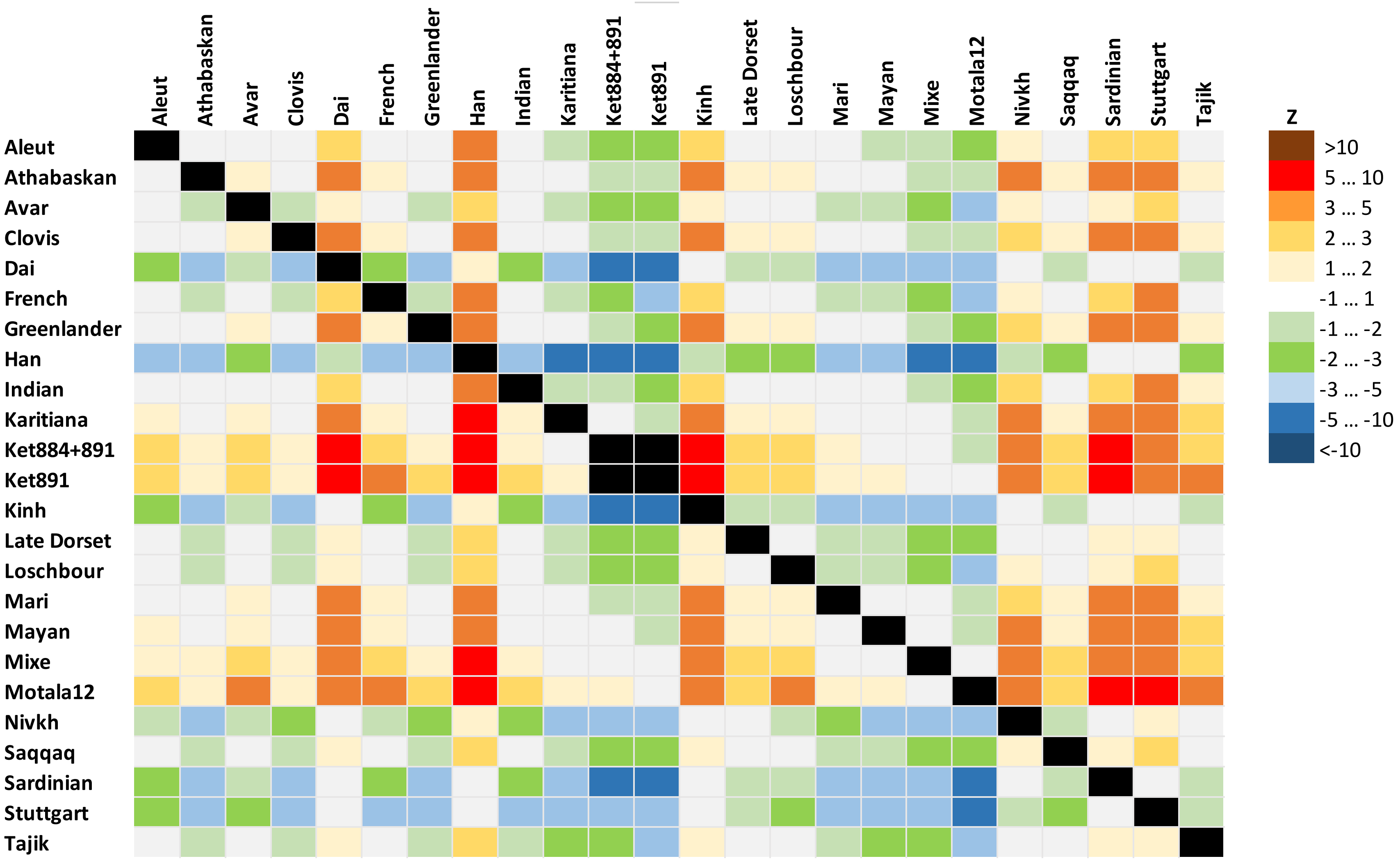

Fig. 3, B.

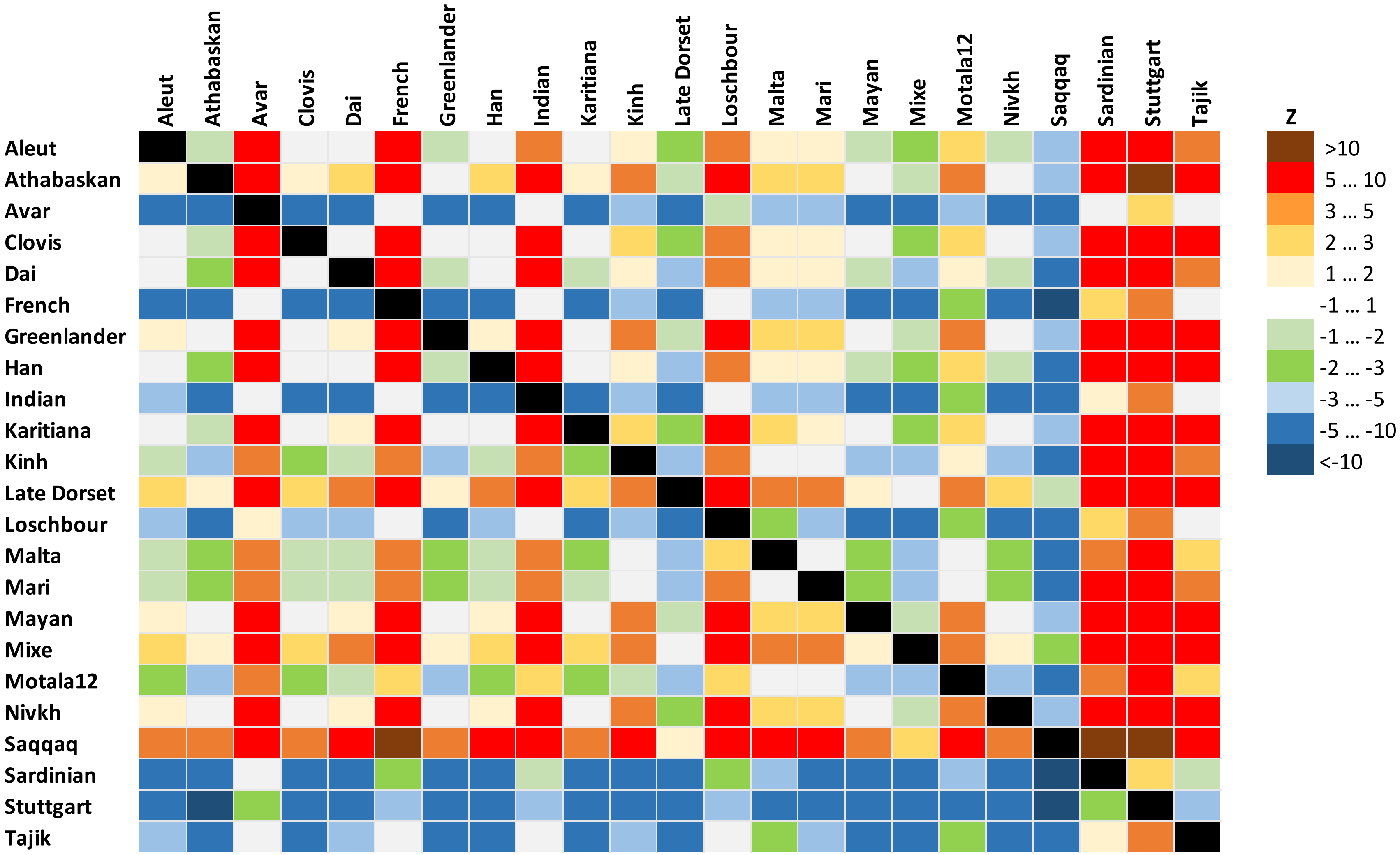

**Fig. 3, C.**

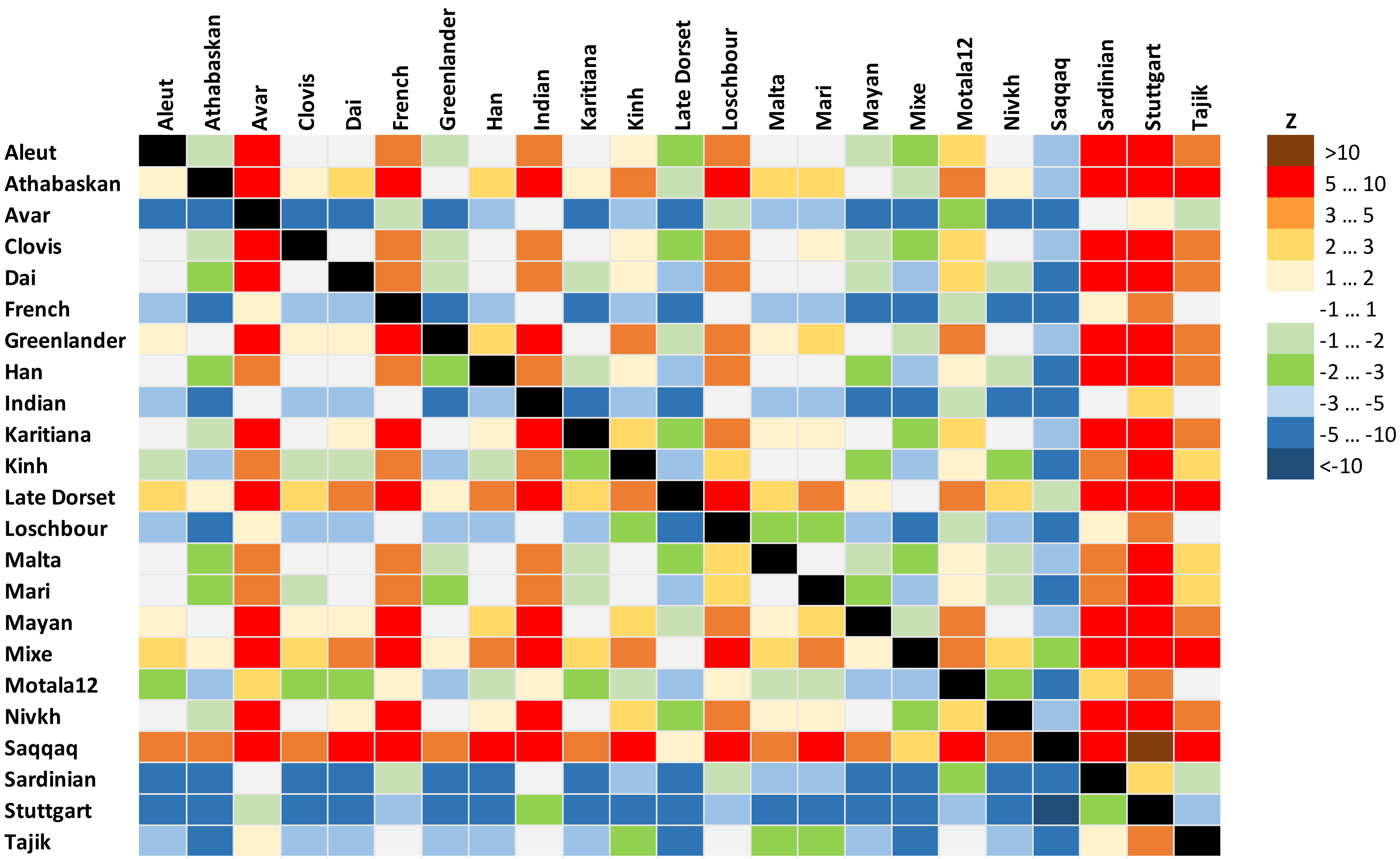

**Fig. 4, A.**

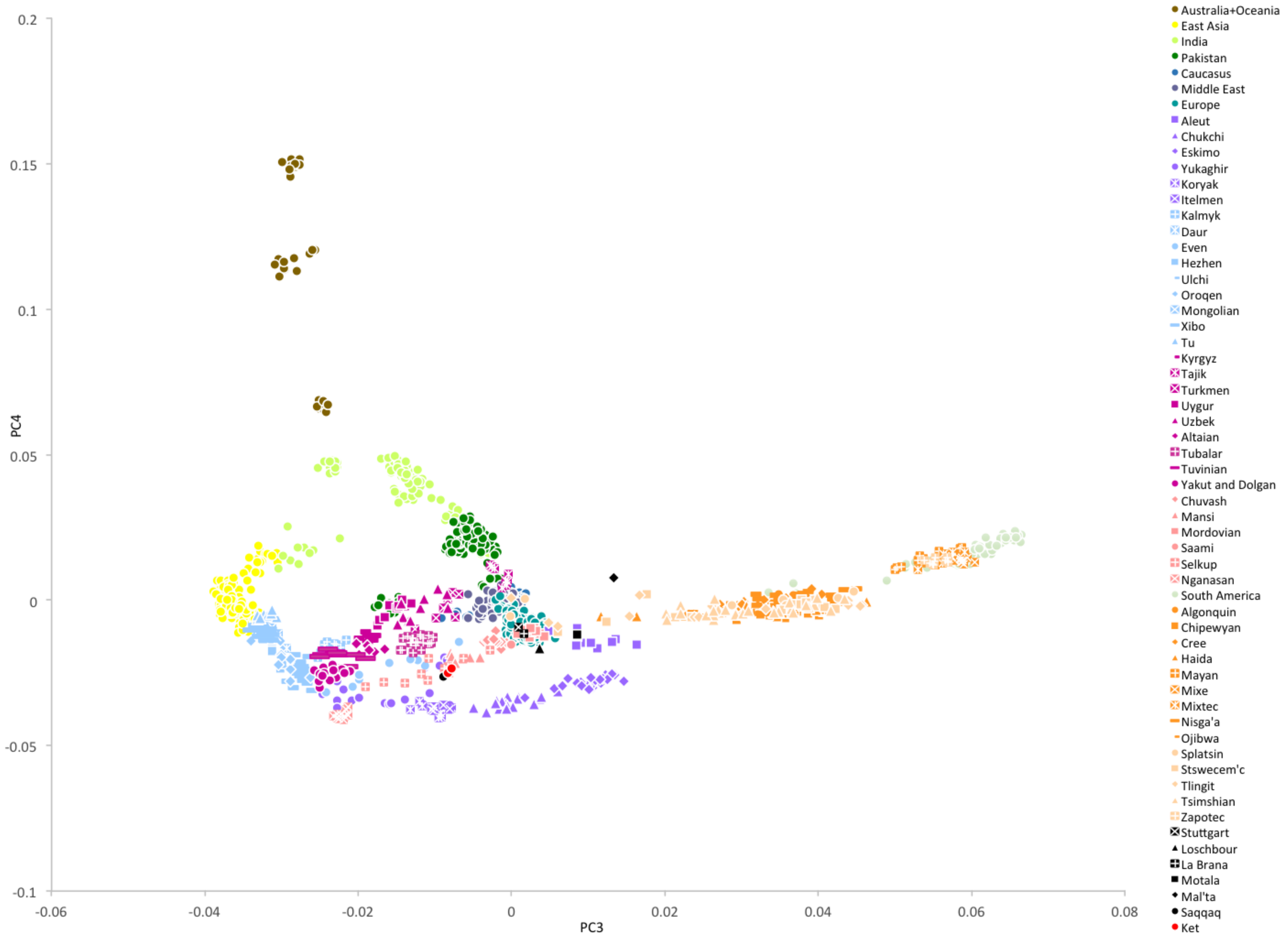

**Fig. 4, B.**

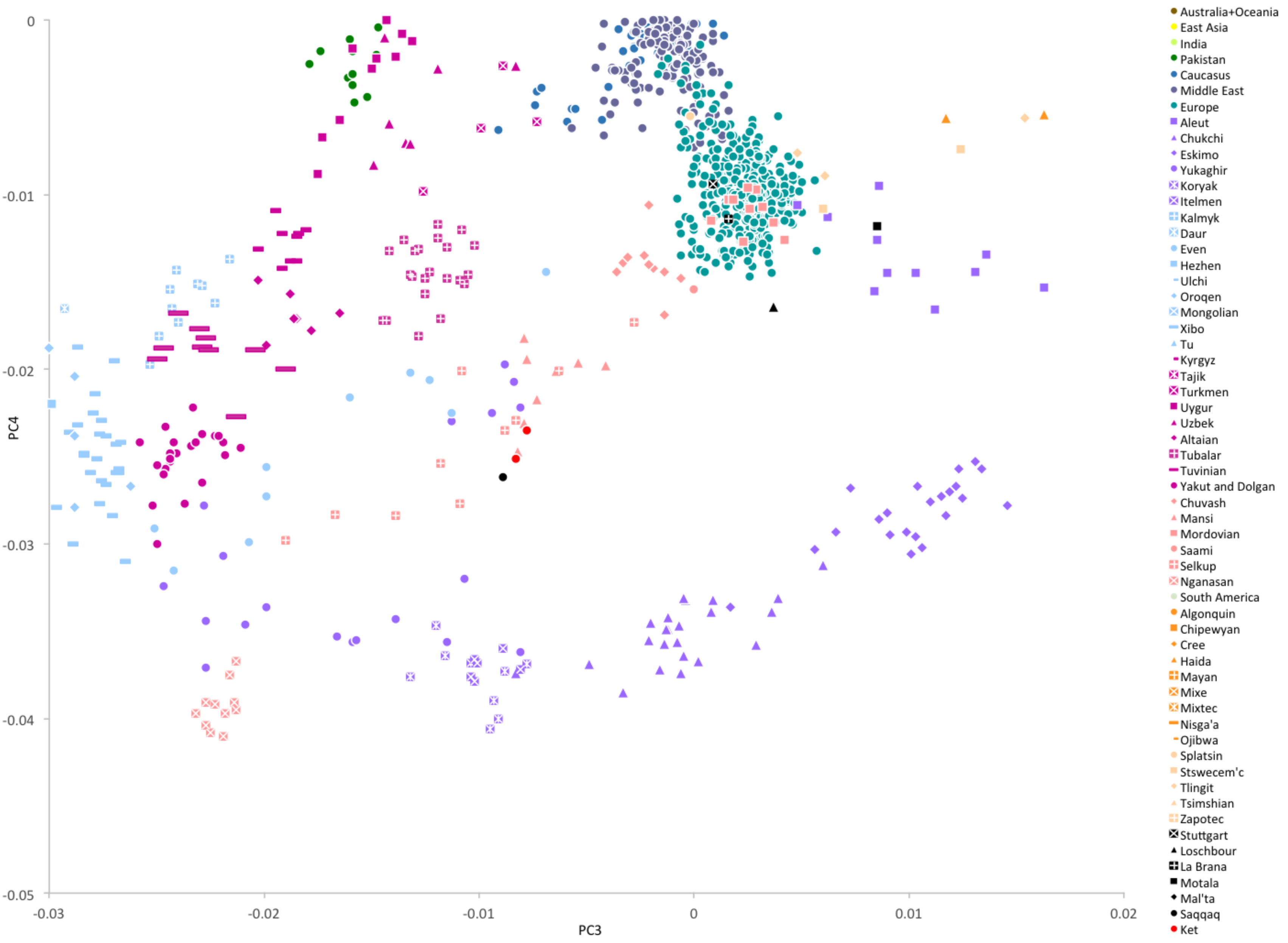

**Fig. 5, A.**

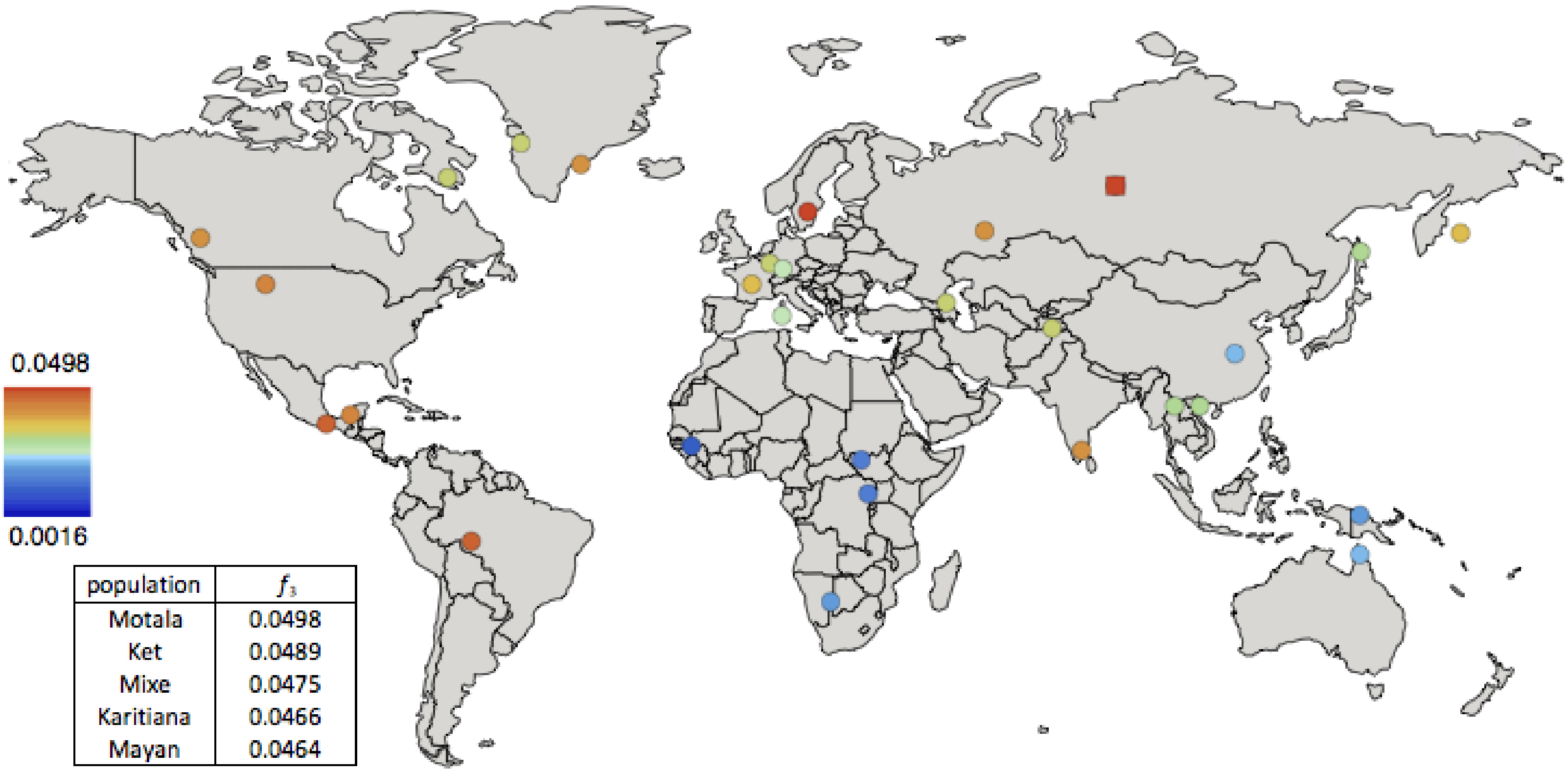

| population | $f_3$ |
|---|---|
| Motala | 0.0498 |
| Ket | 0.0489 |
| Mixe | 0.0475 |
| Karitiana | 0.0466 |
| Mayan | 0.0464 |

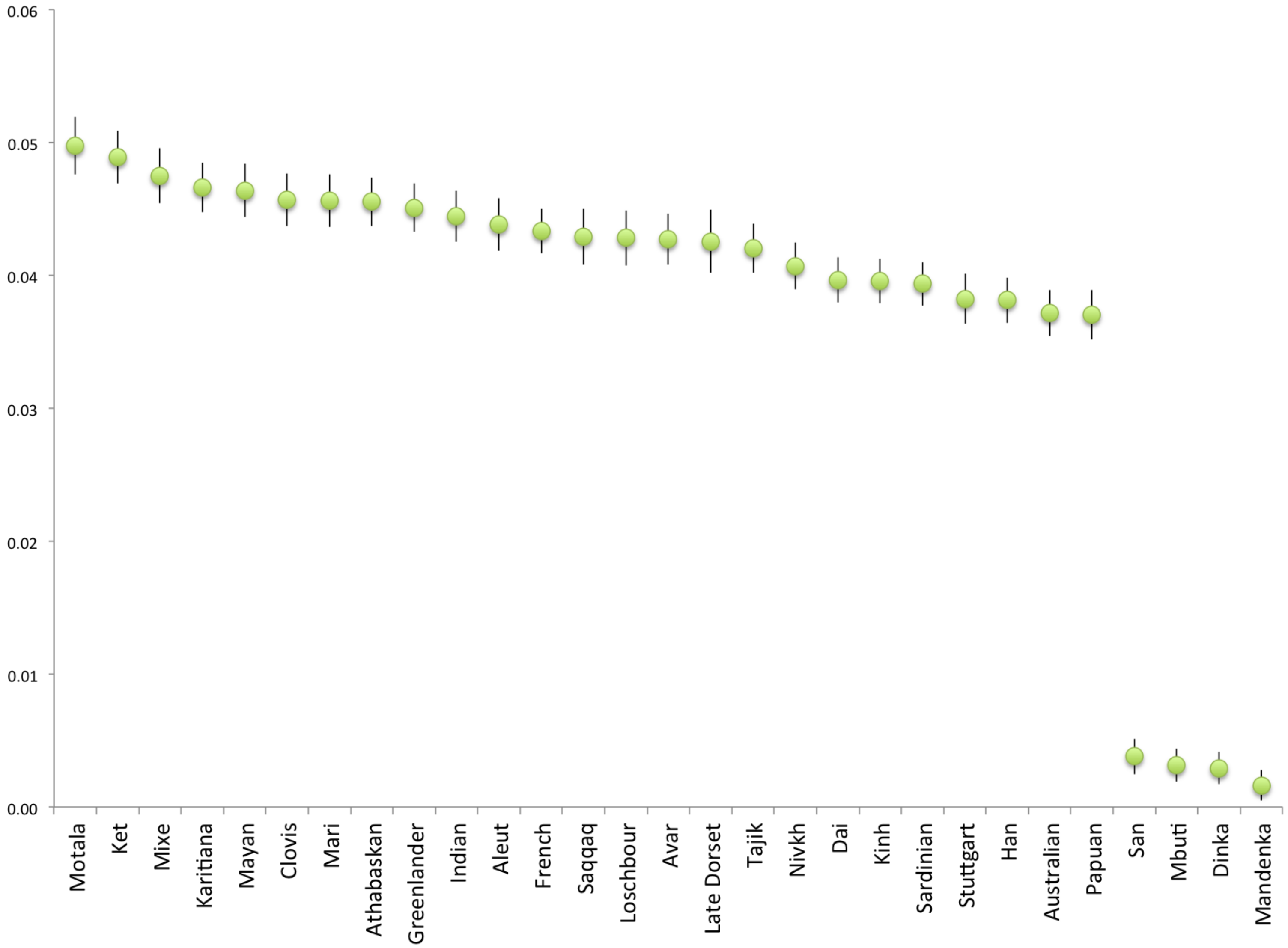

**Fig. 5, B.**

**Fig. 6, A.**

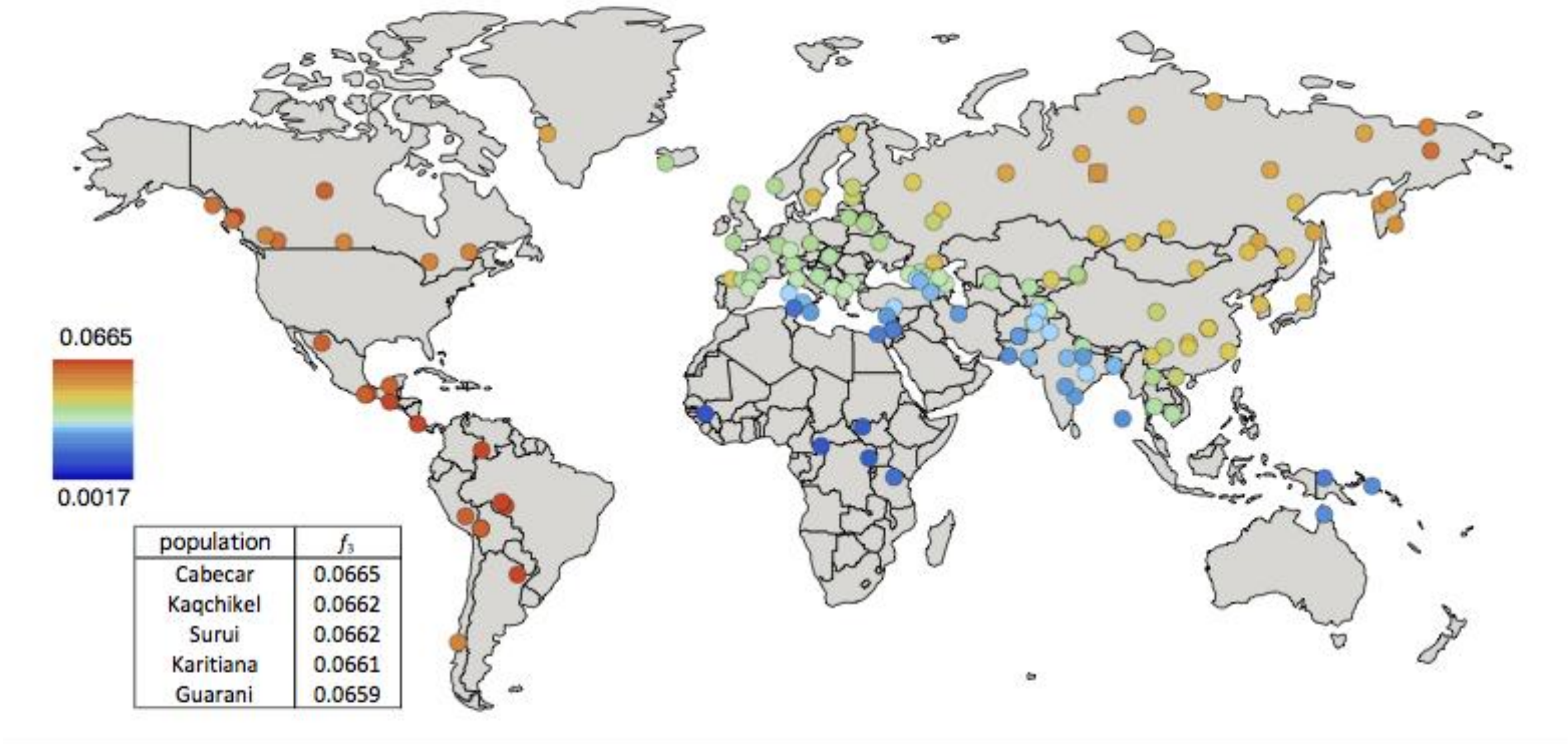

**Fig. 6, B.**

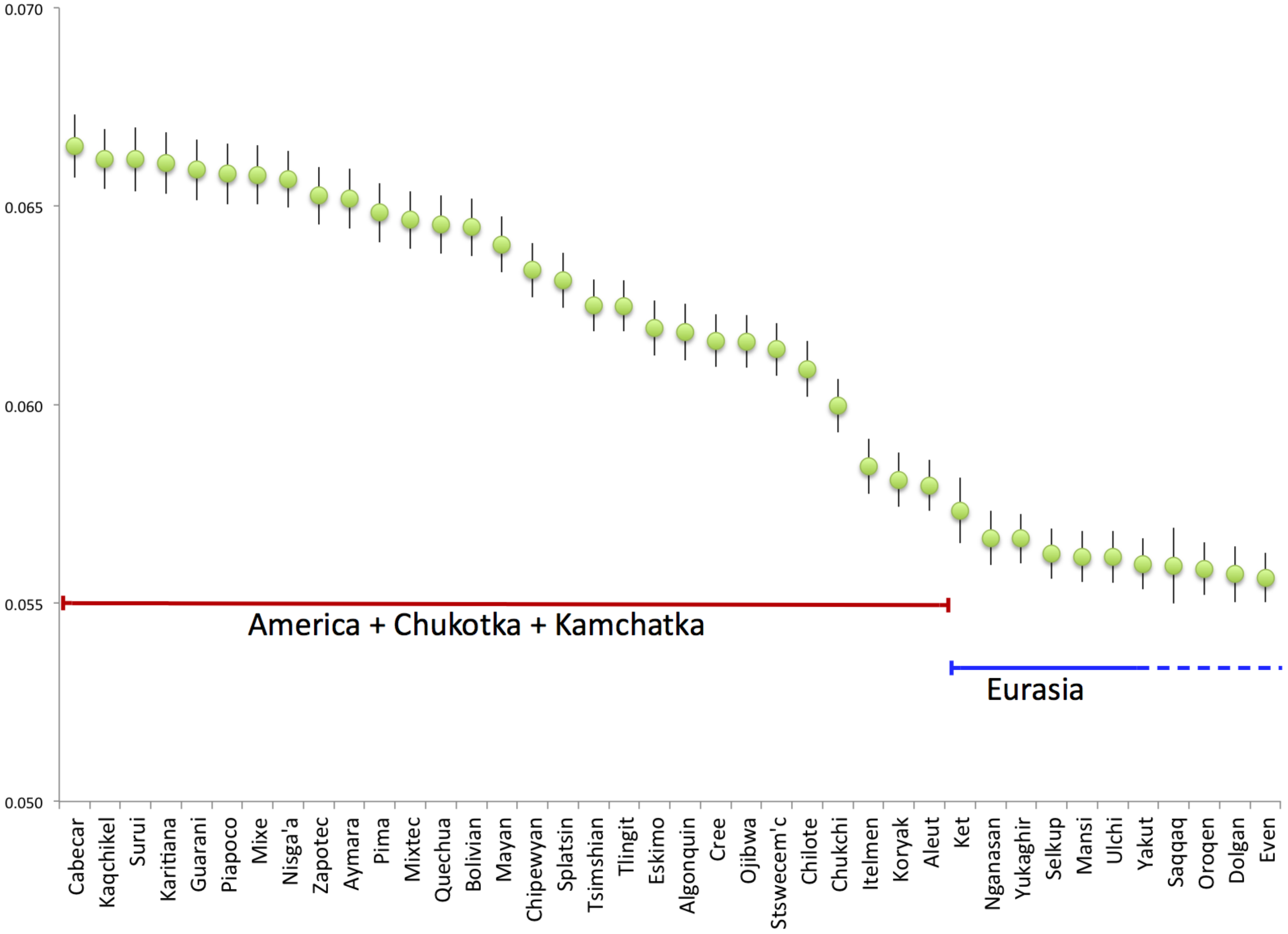

**Fig. 6, C.**

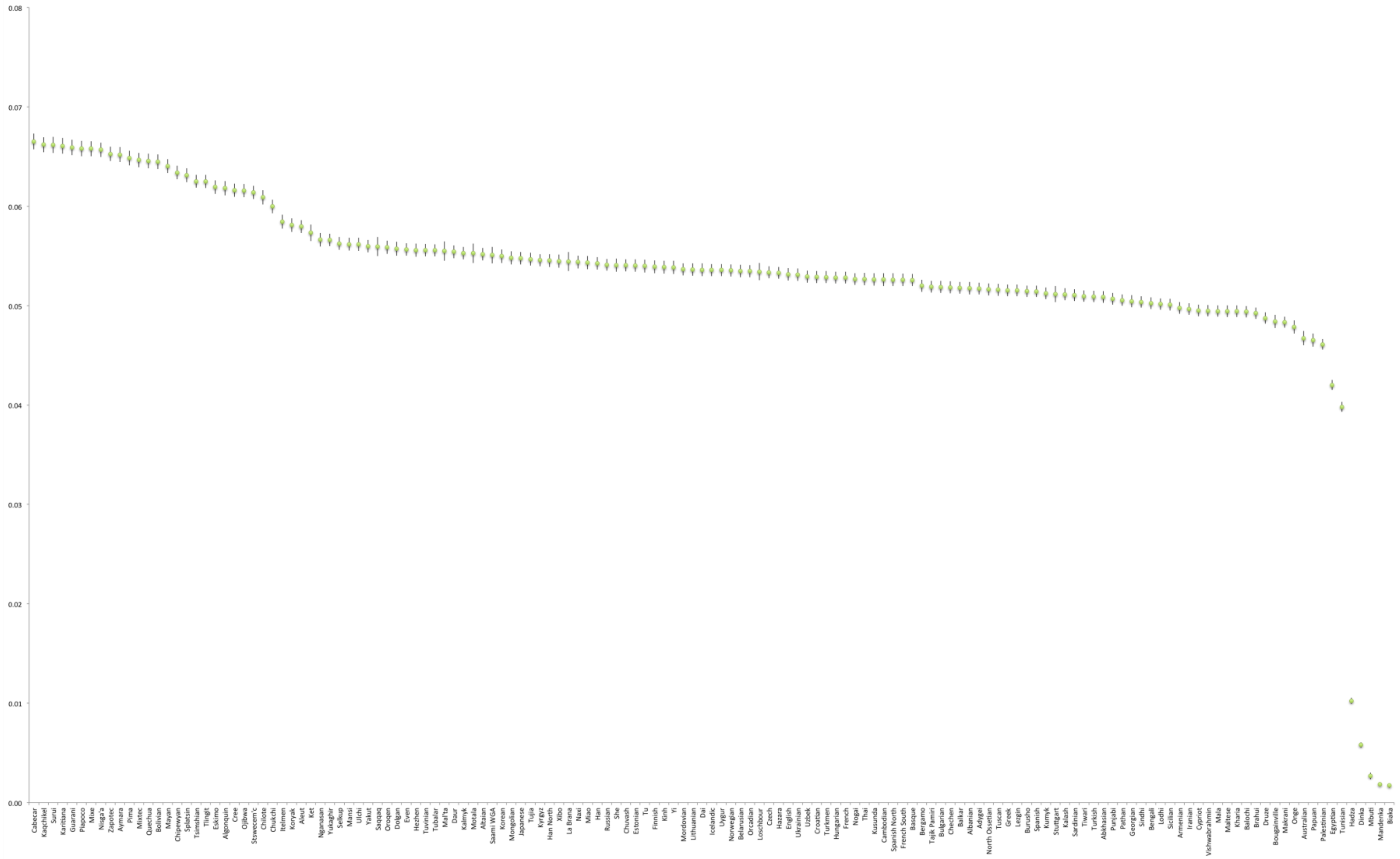